\documentclass[a4paper,twocolumn,11pt]{quantumarticle}
\pdfoutput=1
\usepackage[utf8]{inputenc}
\usepackage[english]{babel}
\usepackage[T1]{fontenc}
\usepackage{amsmath}
\usepackage{hyperref}

\usepackage{tikz}
\usepackage{lipsum}

\usepackage{siunitx}
\usepackage{braket}
\usepackage{amssymb}

\newcommand{\cmmnt}[1]{}

\makeatletter
\providecommand\UseOneTimeHook[1]{}
\makeatother

\begin{document}
\title{Nonlinear Phase Gates Beyond the Lamb-Dicke Regime
}

\author{Akram Kasri}
\affiliation{Palacký University, 77900 Olomouc, Czech Republic}
\orcid{0000-0002-2445-2701}
\email{akramabderrahmane.kasri@upol.cz}
\author{Kimin Park}
\email{park@optics.upol.cz}
\orcid{0000-0003-0290-4698}
\affiliation{Palacký University, 77900 Olomouc, Czech Republic}
\author{Radim Filip}
\affiliation{Palacký University, 77900 Olomouc, Czech Republic}
\email{filip@optics.upol.cz}
\orcid{0000-0003-0290-4698}

\maketitle

\begin{abstract}
Nonlinear phase gates are essential to achieve the universality of continuous-variable quantum processing and its applications. We present a deterministic protocol for generating nonlinear phase gates in trapped ion systems  using simultaneous two-tone sideband drives beyond the Lamb-Dicke regime. Our approach harnesses higher-order interaction terms  typically neglected or suppressed to construct nonlinear phase gates. This methodology enables high-fidelity gate engineering with a near three-fold reduction in  control pulses  compared to  state-of-the-art theoretical proposals.
\end{abstract}

Trapped-ion systems are a leading platform for quantum computation and simulation \cite{leibfried2003,Hffner2008,blatt2012quantum,Katz2023PhysRevLett.131.033604Simulation}, offering excellent qubit coherence and high-fidelity motional mode control \cite{FluhmannNature2019,GanPhysRevLett2020ConditionalBeamSplitter, deNeeve2022}. 
Internal electronic states form high-quality qubits, and  their interactions with  quantized motional modes provide a powerful resource for bosonic quantum information processing \cite{CiracZoller1995PhysRevLett.74.4091TrappedIons,BallanceLucas2016PhysRevLett.117.060504Hyperfine, FluhmannNature2019,Matsos2024}. 
Historically, gate operations  rely on coupling the ion's internal states to its motion via laser-driven sideband transitions, often operating within the Lamb-Dicke (LD) regime  to simplify the dynamics \cite{Park2024NPJQIEfficientCubic}, {and actively suppressing higher-order sideband terms ($k \ge 2$) typically viewed as error sources \cite{Roos2008}}.
{Methods for engineering nonlinear Hamiltonians include} combining multiple linear spin-motion interactions to generate effective nonlinear terms through  series expansion~\cite{SutherlandPhysRevA2021UniversalHybrid, Park2024NPJQIEfficientCubic}. 
In this regime, the requirement for low motional occupation and consequent low gate efficiency limit the complexity of  bosonic nonlinear gate operations, hindering protocols for continuous-variable quantum computation~\cite{BraunsteinRMP2005, WeedbrookRMP2012}.
In this LD limit, however, higher-order interaction terms are neglected, discarding potentially rich dynamics and {increasing circuit complexity}  \cite{AedoPhysRevA2018GeneralDickeSimulation,SametiPRA2021}. 
Extending operations beyond the LD limit introduces significant complexity in designing the required control fields as a cost,  often requiring numerical techniques like robust quantum optimal control  \cite{LiuCPL2025FastGatesBeyondLDaOptimalControl}, and thus has primarily been applied to state preparation \cite{WeiPhysRevA2004BeyondLD}. 
{Early control studies utilized the zeros of Laguerre polynomials beyond the LD limit to truncate Hilbert spaces for better controllability \cite{Rangan}.} 

{Unlike previous methods relying on sequences of linear operations  \cite{Bazavan2024squeezing}, probabilistic subtraction \cite{Madsen2022, Konno2024} or  activating native nonlinearities in hardware with residual Kerr terms \cite{Eriksson2024universal,Sivak2023}, in this work we use two-tone pulses to harness  higher-order terms.}
This approach enables the deterministic and efficient synthesis of nonlinear phase gates, crucial for continuous-variable quantum computation {and simulation} \cite{BudingerPRR2024cubiccomputing,anteneh2025deepreinforcementlearningneardeterministic}, by a sequence of multi-sideband transitions.
By turning parasitic effects into a {resource}, our method provides a more direct path to constructing complex quantum gates.


Our objective is to deterministically synthesize the nonlinear phase gates, defined by the unitary operation $\mathcal{U}^{(j)}=e^{i \zeta_j \hat{X}^j}$ or equivalently $e^{i \zeta_j \hat{P}^j}$ for integer orders $j\ge 3$, where $\hat{X} = (\hat{a} + \hat{a}^\dagger)/\sqrt{2}$ and $\hat{P} = (\hat{a} - \hat{a}^\dagger)/\sqrt{2}i$ are the  quadratures and $\zeta_k$ is the {nonlinearity} strength. Our primary target is the lowest order gates, e.g. $j=3$ (cubic) and $4$ (quartic). 
{We define $\zeta_3$ as the target cubicity parameter for the cubic phase gate.}
The interaction between a trapped ion’s internal qubit states and its external motional degrees of freedom {follows the} interaction Hamiltonian~\cite{leibfried2003}:  
\begin{equation}
   \resizebox{0.425\textwidth}{!}{$
    H = f(t) \hat{\sigma}^+ \exp \left( i \eta \left( \hat{a} e^{-i \nu t} + \hat{a}^\dagger e^{i \nu t} \right) \right) + \text{H.c.}
    $},
    \label{eq:LDE}
\end{equation}
where \( \hat{\sigma}^+ = |e\rangle\langle g| \) is the qubit raising operator, \( \hat{a} \) and \( \hat{a}^\dagger \) are the annihilation and creation operators for the motional mode, \( \eta \) is the LD parameter, $\nu$ is motional mode frequency and \( f(t) \) represents the time-dependent {complex} driving field at time $t$. 
Expanding the exponential term leads to the decomposition into the sideband interaction terms:  
\begin{equation}
    \resizebox{0.425\textwidth}{!}{$\exp \left( i \eta (\hat{a} e^{-i \nu t} + \hat{a}^\dagger e^{i \nu t} ) \right) 
    = e^{-\eta^2 / 2} \sum_{k=-\infty}^{\infty} \mathcal{D}_k(\eta) e^{i k \nu t}$}
    \label{eq:expanding}
\end{equation}
where the operators \( \mathcal{D}_k(\eta) \)  define the coupling to the $k$-th motional sideband, are:  
\begin{equation}
    \mathcal{D}_k(\eta) = \sum_{n=0}^{\infty} \frac{(i \eta)^{2n+k}}{(n+k)!} \hat{a}^{\dagger (n+k)} \hat{a}^n.
    \label{eq:DK}
\end{equation}

We apply a two-tone pulse sequence,  simultaneously {driving}   the red and blue resonances for the $k$-th sideband. 
Each pulse is described by a {complex} driving field {defined as}:
\begin{equation}
    f(t) = \Omega_{k} \left(e^{i( \delta_k t + \phi_k)} + (-1)^k e^{-i (\delta_k t + \phi_k)} \right).
    \label{eq:f(t)}
\end{equation}
where \(\Omega_{k} \) is the Rabi frequencies which are fixed, and \( \delta_k \) is the 
detuning, set to the motional sideband frequency $\delta_k = k \nu$  \cite{leibfried2003}. 
The factor $(-1)^k$ ensures the correct symmetry of the drive. 
For the cubic-phase gate, $\phi_k$ does not require optimization, 
as fixed discrete values $\phi_1 = 0$ and $\phi_3 = \pi$ ensures the correct relative sign  for the effective basis rotation. 
{Physically, this phase configuration aligns the interference between the first and third sidebands to  transform} the natural momentum-like dynamics of the Hamiltonian ({e.g., as in} Eq.~(\ref{eq:H3})) into the target position-dependent gate. 

In contrast, for higher-order gates such as the quartic phase gate {discussed later}, $\phi_k$ is treated as a free parameter to account for the additional phase relationships between multiple sidebands.

{This time-dependent pulse simplifies the evolution via a time-independent Hamiltonian.}
Specifically, by moving to an interaction picture rotating at the sideband frequency $\delta_k$ and applying the rotating-wave approximation, the interaction Hamiltonian from Eq. (\ref{eq:LDE}) simplifies to \cite{SametiPRA2021}:
\begin{equation}
    \resizebox{0.425\textwidth}{!}{$
    H_{1} = \frac{i}{4} \Omega_{1} \hat{\sigma}_y \left[ \eta (\hat{a}^\dagger - \hat{a}) + \frac{\eta^3}{2} (\hat{a}^\dagger\hat{a}^2 -\hat{a}^{\dagger 2} \hat{a}   )   \cdots  \right]$.}
    \label{eq:H1}
\end{equation}
{Similarly, driving the second sideband ($\delta_2 = 2\nu$) generates the squeezing interaction. 
In the interaction picture, the expansion yields:
\begin{equation}
\resizebox{0.425\textwidth}{!}{$
H_{2} = \frac{-i}{4}\Omega_{2}\hat{\sigma}_y \left[ \frac{\eta^2}{2!} (\hat{a}^{\dagger 2} + \hat{a}^2) - \frac{\eta^4}{3!} (\hat{a}^{\dagger 3}\hat{a} + \hat{a}^{\dagger}\hat{a}^3) + \dots \right].
\label{eq:H2}
$}
\end{equation}

The third sideband generates:}
\begin{equation}
    \resizebox{0.425\textwidth}{!}{$
    H_{3} = \frac{-i}{4} \Omega_{3} \hat{\sigma}_y \left[ \frac{\eta^3}{3!}  (\hat{a}^{\dagger 3} - \hat{a}^3) + \frac{\eta^5}{4!} (-\hat{a}^\dagger\hat{a}^4  + \hat{a}^{\dagger 4} \hat{a}) \cdots\right].
    $}
    \label{eq:H3}
\end{equation}
{Rabi frequencies $\Omega_{k}$ are assumed to have the same value, and LD parameter $\eta$ is fixed.}

{Hamiltonians $H_{k}$ provide the gate primitives in the LD limit ($\eta \ll 1$): $H_{1}$ in (\ref{eq:H1}) induces displacement $\alpha \propto \eta \Omega_{1} t_1$, $H_{2}$ in (\ref{eq:H2}) provides squeezing $r \propto \eta^2 \Omega_{2} t_2$, and $H_{3}$ in (\ref{eq:H3}) generates triplicity $\chi_{tri} = \frac{i}{4!} \Omega_{3} \eta^3 t_3$ \cite{Bazavan2024squeezing}, {approximately identifying the target cubicity $\zeta_3 \approx \chi_{tri}$}. 
Beyond the LD limit, the $\mathcal{O}(\eta^3)$ term of $H_{1}$ provides a number-dependent momentum shift essential for curving the state into a crescent, while $\mathcal{O}(\eta^4)$ terms in $H_{2}$ are harnessed by the optimizer to compensate for residual distortions from $H_{1}$ and $H_{3}$.}
 However, operating beyond the LD regime introduces a trade-off. The higher order terms beyond the leading first terms in Eqs. \textcolor{black}{(\ref{eq:H2}-\ref{eq:H3}) and beyond the second term in (\ref{eq:H1})} act as nonlinear distortions other than for $U'_1$. In our protocol, these are treated as a cost that the global optimizer must compensate for to maintain high fidelity.

\begin{figure*}[t!]
\centering
\includegraphics[width=0.9\textwidth]{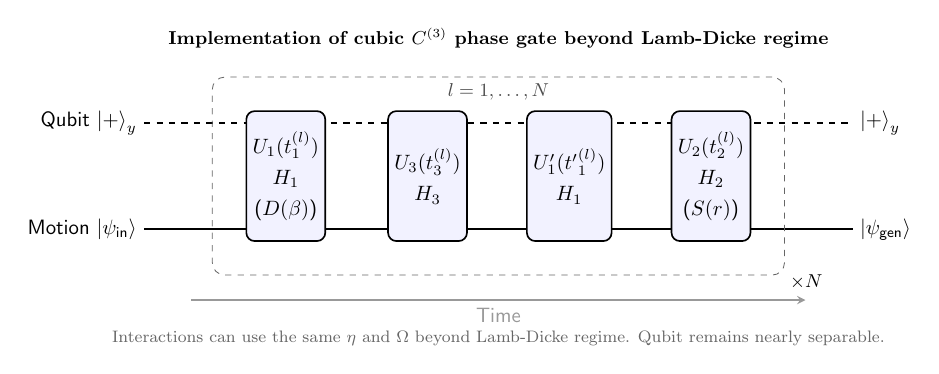}
\caption{Quantum circuit for the cubic phase gate protocol in Eq.~(\ref{eq:protocol}). The protocol consists of repeated application of the composite gate sequence $U_1$, $U_3$, $U_1$, $U_2$ applied $N$ times, {optimized over all $t_k^{(l)}$  for fixed $\eta$ and $\Omega_{k}$}. 
{While each unitary $U_k$ is realized via multi-tone sideband pulses beyond the Lamb-Dicke regime,} {$U_1$ and $U_2$ can be alternatively 
implemented as high-fidelity laser-free operations: displacement $D(\beta)$ via resonant classical 
electric fields \cite{LoNature2015SpinDependentforce} and squeezing $S(r)$ via parametric trap modulation \cite{burd2019quantum, burd2024squeezing}.}
Since the qubit occupies the $|+\rangle_y$ eigenstate of the of $\hat{\sigma}_y$ and interaction Hamiltonians, it remains stationary throughout the evolution. 
}
\label{fig:circuit}
\end{figure*}

Building on these controlled interactions, our deterministic protocol shown in Fig. \ref{fig:circuit} is 
\begin{equation}
 C_{N}^{(\mathbf{3})}(t) = \prod _{l=1}^N U_2^{(l)}{U'}_{1}^{(l)} U_{3}^{(l)}U_1^{(l)},
  \label{eq:protocol}
\end{equation}
where $l$ is the round index, that acts on the joint qubit-motional state 
$|\psi_\mathrm{in}\rangle\!\otimes\!(|e\rangle\!\pm\!  i\,|g\rangle)$ as follows, where $|\psi_\mathrm{in}\rangle$ is the motional initial state.
{The unitaries are defined as $U_j^{(l)}=\exp[i t_j^{(l)} H_j]$, where $H_j$ are Hamiltonians in (\ref{eq:H1}, \ref{eq:H2}, \ref{eq:H3}). All $t_1^{(l)}$, $t_3^{(l)}$, ${t'}_1^{(l)}$, $t_2^{(l)}$,  are times for individual unitary operators optimized for fixed $\Omega_{k}$ and $\eta$.
The total time $t$ is therefore given as $t=\sum_l (t_1^{(l)}+t_2^{(l)}+t_3^{(l)}+{t'}_1^{(l)})$.}
{$\zeta_k^{\text{eff}}$ denotes the effective strength realized by the protocol.}
Our protocol utilizes the qubit as an ancilla, which \textcolor{black}{remains} disentangled from the motional state obtaining effectively same qubit state throughout the gate sequence, ensuring that the net operation acts solely on the oscillator deterministically. We also explored several alternative orderings of the  unitaries \(U_1\), \(U_2\) and \(U_3\) within each  round in the circuit in Fig. \ref{fig:circuit} and found that, even after re-optimizing the {protocol with the lowest} $N=1$, the fidelity saturated around $F \simeq 0.96$, well below the $F \simeq 0.998$ achieved with the ordering used in our protocol.

{We consider two physical implementations for the displacement $U_1$. First, using the sideband interaction $H_{1}$, a pure displacement is recovered only in the strict Lamb-Dicke limit ($\eta \ll 1$). Since the displacement amplitude scales as $\beta \propto \eta \Omega t$, maintaining a constant $\beta$ while reducing $\eta$ requires a proportional increase in pulse duration ($t_1 \propto 1/\eta$), leading to {a long} time requirement {for a small} $\eta$. Our protocol avoids this by operating at $\eta \approx 0.3$ beyond LD regime and utilizing the optimizer to manage the resulting nonlinear distortions. 
Alternatively, {displacement} can be implemented via an auxiliary resonant classical drive (direct electric field) \cite{LoNature2015SpinDependentforce}. This interaction is fundamentally $\eta$-independent as it relies on a long-wavelength field, yielding a state-independent displacement {without nonlinear terms and with a speed increased by field amplitude}.  
While this classical drive provides a {precise}, fast displacement, the sideband approach allows for a unified hardware control sequence {without auxiliary tools} where the qubit acts as a mediating ancilla throughout all gate components.}
 
The unitary $U_3$ is applied {sequentially {after} $U_1$} {with fixed $\eta$ and $\Omega_{3}$},  introducing the core third-order terms of the target {cubic phase gate $\exp[i \zeta_3 \hat{P}^3]$}.
The residual terms primarily consist of {uncompensated $\mathcal{O}(\eta^4)$ distortions from Eqs.~(\ref{eq:H1}-\ref{eq:H3}) and $\mathcal{O}(\eta^5)$ terms from the full expansion of Eq.~(\ref{eq:LDE})}.

The \cmmnt{physical} interaction $H_{1}$ now contribute the crucial nonlinear terms $i (\hat{a}^{2}\hat{a}^{\dagger} - \hat{a}^{\dagger 2}\hat{a})$. 
Physically, this term acts as a number-dependent momentum transformation ($\propto \hat{n}\hat{P}$) that induces a self-shift of a thermal state along the position quadrature $\hat{X}$, with a rate proportional to the mean phonon number. This nonlinear behavior is essential for producing the non-Gaussian character of the target cubic-phase operator $\hat{X}^3$.

Crucially, the dominant first-order linear term $\eta(\hat{a}^\dagger - \hat{a})$ in $H_{1}$ in (\ref{eq:H1}) is managed by the balanced application of $U_1$ and $U_1'$. While $U_1$ provides a necessary displacement that shifts the motional state into a regime where nonlinearities are more pronounced, the subsequent $U_1'$ is optimized to counteract this shift. This ensures that the linear displacements cancel out over each round $l$, leaving only the accumulated nonlinear phase. 

Because the first-sideband Hamiltonian $H_{1} \propto \hat{\sigma}_y$, 
its {nonlinear} contribution must {interfere constructively with} $H_{3}$ for faithful gate implementation. 
Formally, this corresponds to introducing a flip in the qubit subspace ($\hat{\sigma}_y \to -\hat{\sigma}_y$), 
which in principle could be realized by an explicit qubit $\pi$-pulse (e.g., $R_x(\pi)$) without affecting the oscillator. 
In our implementation, however, {the sign reversal is implemented without auxiliary gates.} 
The same effect is obtained directly through the phase parameter $\phi_k$ already defined in the driving field $f(t)$. Changing the relative phase between the two drives by $\pi$ modifies the interaction term in the rotating frame as 
$f(t) \rightarrow -f(t)$, which in turn reverses the sign of the effective coupling proportional to $\hat{\sigma}_y$ \cite{lee2005phase}.

Finally, the 
operation $U_2$ is deterministically generated using  {$H_{2}$ in (\ref{eq:H2})}
acting as a {qubit-mediated squeezer} \cite{solano2002entangled, Jeon2024EntangledCoherent}. 
While  higher-order parasitic terms in Eq. (\ref{eq:H2}) of $\mathcal{O}(\eta^4)$ introduce number-dependent squeezing distortions, their magnitude remains small ($\approx \eta^2/3$ relative to the leading {squeezing term $\frac{\eta^2}{2}$, which is $\approx 0.045$ for $\eta=0.3$)}). 
{This  high-fidelity squeezing  can be generated via direct 
parametric modulation of the trap electrodes \cite{burd2019quantum,burd2024squeezing}, which can potentially reduce the number of sideband pulses. This mechanical parametric drive with squeezing levels of up to $20.6 \pm 0.3$ dB allows for required pure squeezing of the motional mode even when the ion is in a large-extent state.}
The global optimizer specifically tunes $t_2$ to achieve the target squeezing amplitude $r$ in the squeezing $S(r) = \exp[r(\hat{a}^2 - \hat{a}^{\dagger 2})/2]$ while ensuring these residual nonlinearities are constructively compensated for by the $U_1$ and $U_3$ sequences.
Within LD approximation, this squeezing operation  can be implemented with high fidelity using a resonant pulse, although long pulse durations may introduce decoherence where \(|r|= \frac{1}{8}\Omega_{2}\eta^2t\) \cite{solano2002entangled}.
Our {deterministic} protocol in Fig. \ref{fig:circuit} allows for the sequential application of {$C_{N}^{(\mathbf{3})}(t)$}. 
Repeating the {circuit} with independently optimized parameters {of all $t_k^{(l)}$'s} allows the {cubic} nonlinearity to accumulate effectively.
We can increase {the circuit depth for optimization},  by repetition number $N$.
The overall sequence of qubit-dependent operations impart a net phase on the motional state while the qubit remains in its initial superposition state $|+\rangle_y$. 

\begin{figure}[h]
    \centering
    \includegraphics[width=\columnwidth]{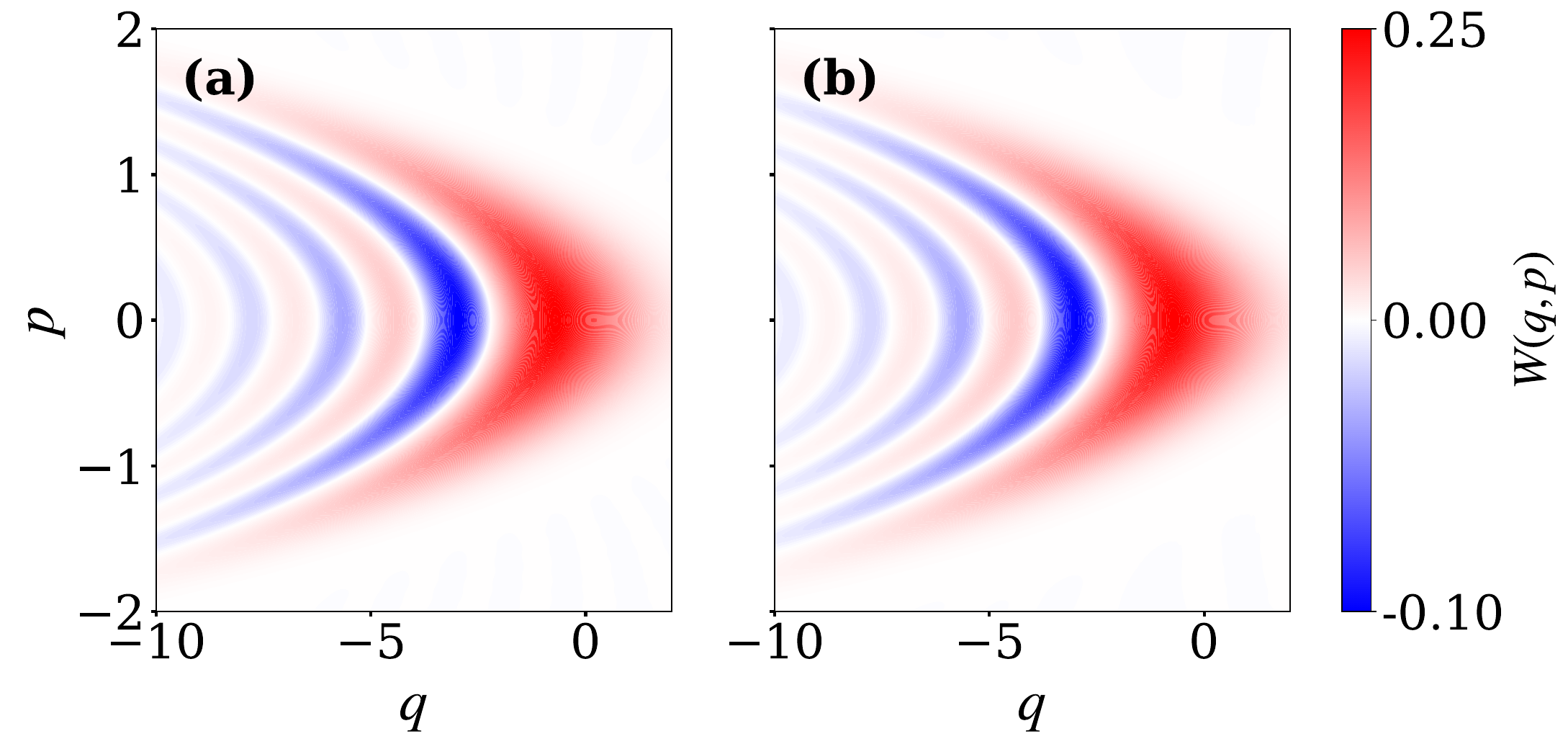}
    \caption{Wigner function representations of (a) the target cubic gate {$\mathcal{U}^{(3)}(\zeta_3)$} with cubicity $\zeta_3= 1$ {on {ground} states as} in (\ref{eq:cubicstate}), and (b) the generated state obtained using the optimized $N=3$, $C_{N=3}^{(3)}$  (parameters in Table \ref{tab:params_3g}). The protocol achieves fidelity $\mathcal{F} = 0.99986$ and closely reproduces the non-classical features, with similar total negativity volumes $V_{-}(\rho) = \frac{1}{2} \int |W_\rho(\mathbf{r})| - W_\rho(\mathbf{r})  d\mathbf{r}$ of $V_{-}(\text{target}) = 0.226$ and $V_{-}(\text{generated}) = 0.2229$. 
    }
    \label{fig:wignercomp}
\end{figure}

\begin{figure*}[t!]
    \centering
    \includegraphics[width=0.8\textwidth]{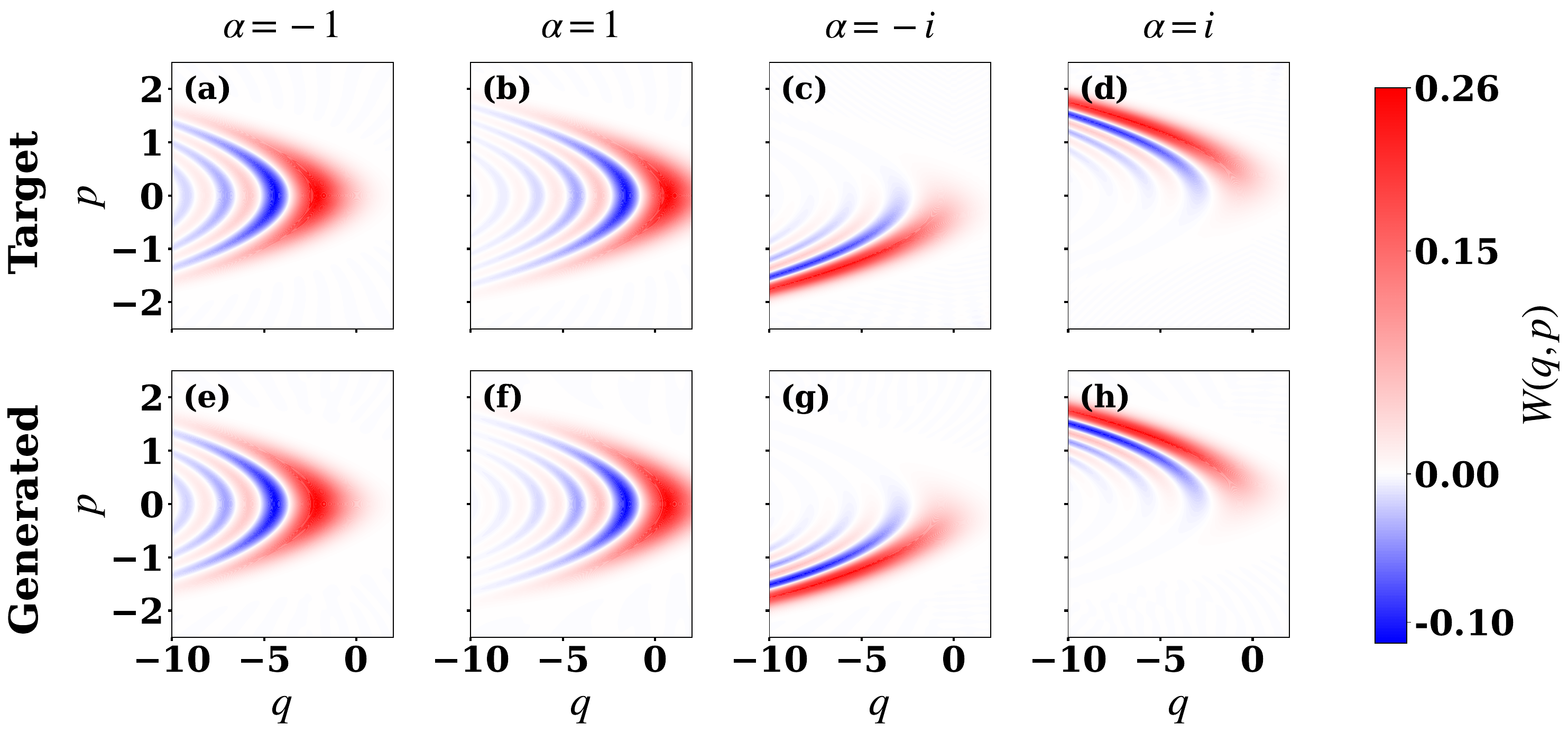}
    \caption{Wigner functions for the cubic gate acting on {coherent state} inputs. Top row: target states $ \mathcal{U}^{(3)}(\zeta_3=1)\,|\alpha\rangle$ for $\alpha\in\{-1,+1,-i,+i\}$. Bottom row: states generated by the optimized {$N=3$} gate {$C_{N=3}^{(3)}$} (same gate parameters as in Table~\ref{tab:params_3g}). Fidelities: $\mathcal{F}(\alpha=1)=0.99937$, $\mathcal{F}(\alpha=-1)=0.99965$, and $\mathcal{F}(\alpha{=}\pm i)=0.969$. These examples complement Fig.~\ref{fig:repetition-fidelity}, where the infidelity trend is reported across purely imaginary $\alpha$.}
 
    \label{fig:wignercoherent}
\end{figure*}

\begin{table}[!h]
\centering
\begin{tabular}{c S S S S}
\hline
{Round} &
\multicolumn{1}{c}{$t'_1$ (\si{\micro\second})} &
\multicolumn{1}{c}{$t_3$ (\si{\micro\second})} &
\multicolumn{1}{c}{$t_2$ (\si{\micro\second})} &
\multicolumn{1}{c}{$\beta$} \\
\hline
1 & 92.35 & 88.75  & 13.6  & -2.84 \\
2 & 77.58 & 119.63 & 83.9  & 2.00 \\
3 & 25.99 & 157.65 & 137.5 & -2.54 \\
\hline
\end{tabular}
\caption{Optimized parameters for the $N=3$ protocol used to generate states in Figures \ref{fig:wignercomp} and \ref{fig:wignercoherent},
with $\eta=0.3$, $\Omega_{k}=0.3$~MHz. The displacement amplitude $\beta \propto \eta \Omega t_1$ using (\ref{eq:H1}), or is implemented by a classical drive \cite{LoNature2015SpinDependentforce}. The corresponding total gate duration is $t_{\mathrm{tot}} \approx 0.797~\mathrm{ms}$, compared to $\approx 0.07~\mathrm{ms}$ for the previous Fourier-synthesis protocol~\cite{Park2024NPJQIEfficientCubic} if the same LD parameter is used, or much longer if we need to suppress errors. The $t_2$ can be translated to squeezing parameters $(r_1,r_2,r_3)=(0.08,4.91,-8.06)\,[\mathrm{dB}]$ for parametric squeezing.}
\label{tab:params_3g}
\end{table}

Following this sequence (\ref{eq:protocol}), the generated state approximates the ideal cubic phase operation, but includes {residual} contributions from higher-order nonlinear terms:
\begin{align}
|\psi_{\text{gen}}\rangle \approx e^{i\zeta_3\hat{X}^3}|\psi_\mathrm{in}\rangle + \text{residual terms}
\label{eq:cubicstate}
\end{align}
where  \( e^{i\zeta_3 \hat{X}^3} \) corresponds to the cubic phase operation. 
{The residual terms primarily consist of uncompensated squeezing and higher-order nonlinear phase contributions (e.g., $\propto \hat{X}^5$) arising from the non-truncated expansion of Eq.~(\ref{eq:LDE}).}

To yield a nearly pure cubic gate, these residual contributions must be minimized. 
We achieve this by {increasing the number of rounds $N$ and performing a global optimization over the independent pulse durations $\{t_k^{(l)}\}$ for each round $l$}, as demonstrated by the improved performance of our $N=2$ and $N=3$ protocols.
An alternative strategy  would be to use precise pulse control within a single gate sequence \cite{SametiPRA2021}.

\begin{figure}[!h]
    \centering
    \includegraphics[width=\columnwidth]{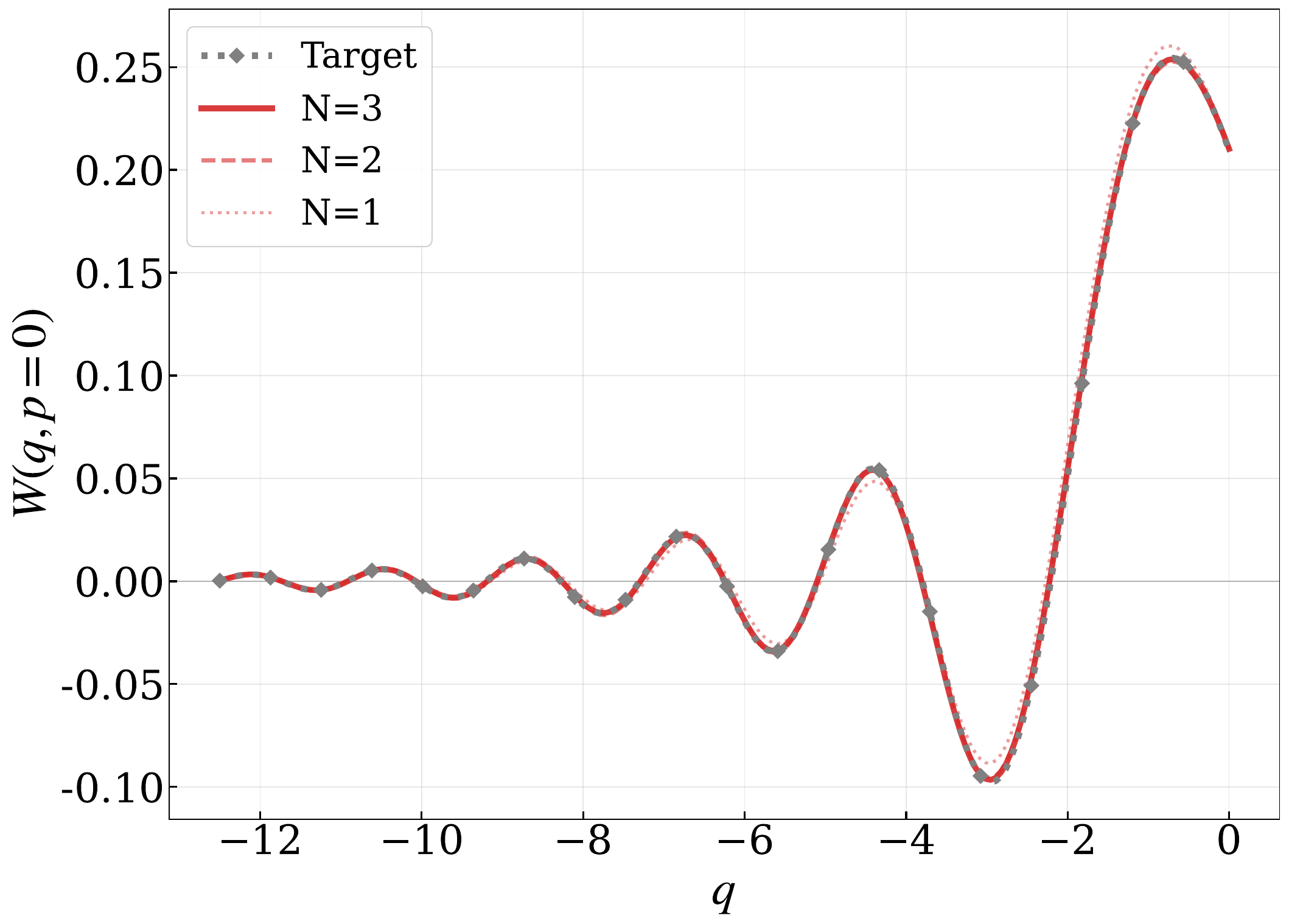}
    \caption{Comparison of Wigner function negativity along {the position axis, i.e.} \(p=0\) for the target cubic phase {gate on the ground state in Fig. \ref{fig:wignercomp}}   and protocols with increasing gate complexity ($N=1, 2, 3$). The {$N=3$} protocol achieves {high} fidelity of 0.99986 with the target state.}
    \label{fig:density_wigner}
\end{figure}

\begin{figure}[t!]
    \centering
    \includegraphics[width=\columnwidth]{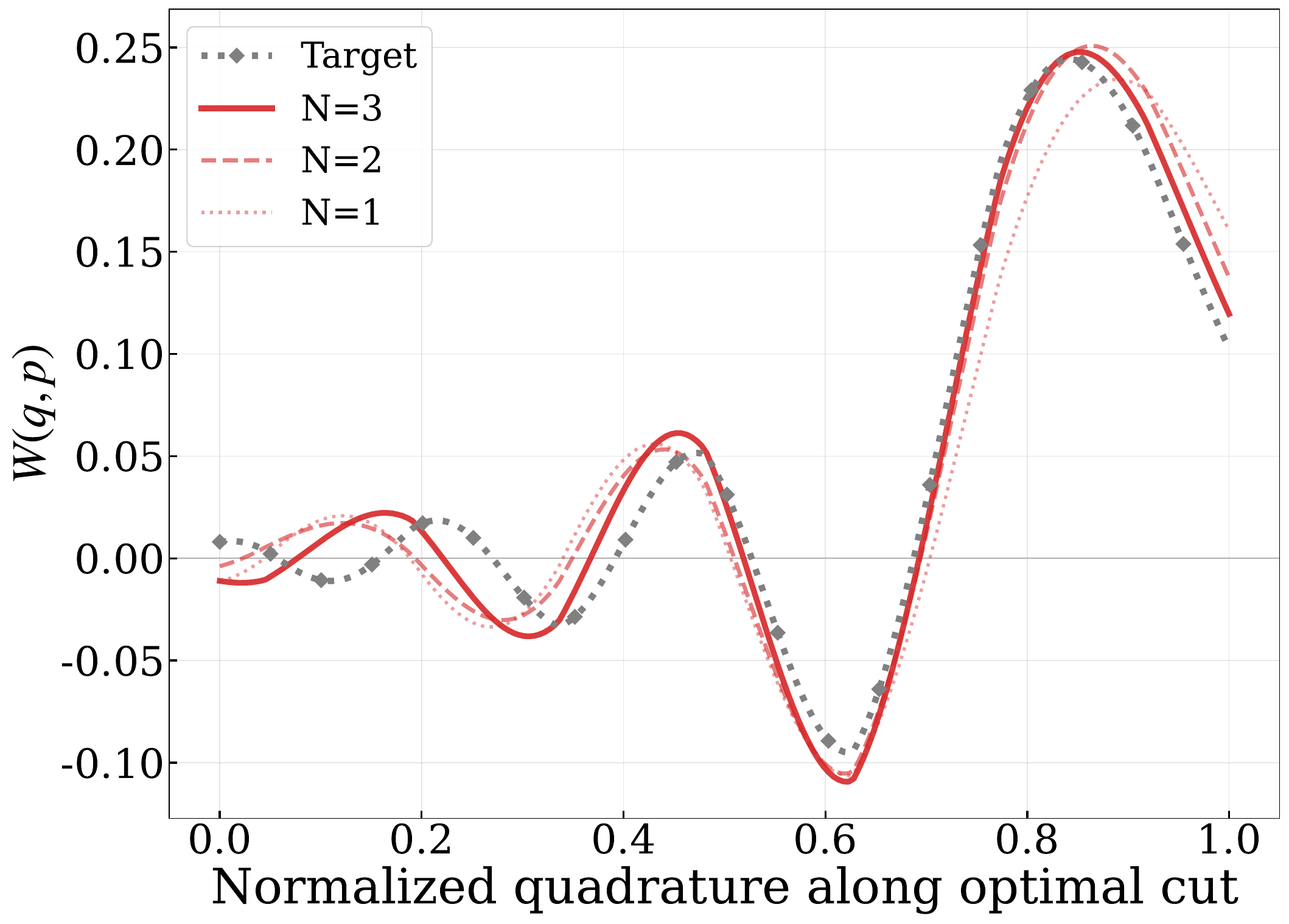}
    \caption{Wigner-negativity cut {from the Figure \ref{fig:wignercoherent}} for the target cubic state and the state generated from the coherent {state} input $\ket{\alpha=-i}$ {by optimized sequence (\ref{eq:protocol})}. The optimal cut follows the line of maximal negativity $p = -0.0687\,q - 1.9703$     between {phase space points} $(q,p)=(-12.50,-1.11)$ and $(-6.22,-1.54)$.}
    \label{fig:diagcut}
\end{figure}

Beyond global metrics like fidelity and variance, Figure~\ref{fig:density_wigner} provides a direct comparison of specific non-classical features of the state by plotting a cross-section of the Wigner function along the {position axis} $p=0$, where its negativity is prominent. {Complementing this, Fig.~\ref{fig:diagcut} illustrates the cross-section along the diagonal of maximal negativity for the {worst case} coherent input $|\alpha=-i\rangle$, confirming that the generated state captures the detailed interference fringes.
These cuts show that the oscillations and negative regions of the generated state's Wigner function progressively approach those of the target state as the gate complexity increases from $N=1$ to $N=3$.}

To accurately approximate the ideal cubic phase gate $e^{i\zeta_3 \hat{X}^3}$, it is essential to balance the nonlinear contributions arising from $H_{1}$ and $H_{3}$. 

While the target operator {$e^{i\zeta_3\hat{X}^3}$} is defined in the position quadrature, the native Hamiltonians generate momentum-like dynamics. As detailed in the theory section, selecting specific $\phi_k$ implements an effective basis rotation, aligning these native interactions with the target operator.
Consequently, the problem reduces to synthesizing the momentum operator $e^{i\zeta_3\hat{P}^3}$.
The fidelity of the generated gate critically depends on {optimization}; any deviation leads to a {distorted} phase profile. Optimal matching of {pulses in Fig. \ref{fig:circuit}} is therefore key to achieving a high-fidelity cubic phase gate. {We also tested a protocol variant in which the first- and third-sideband interactions \( H_{1} \) and \( H_{3} \) are applied simultaneously within a single pulse. 
The total gate duration for this simultaneous scheme is \( t = 167.889 \,\mu\text{s} \), which is less than half the time of the sequential \( N=1 \) protocol $370 \mu\text{s}$. However, the resulting fidelity is slightly lower due to the increased complexity in balancing the non-commuting interaction terms within a single time step, indicating a trade-off between gate fidelity and pulse duration.}

{
Fidelity can be further enhanced by bracketing the protocol with  pre-squeezing $S(r_{\mathrm{pre}})$ independent of the input state $\ket{\psi_\mathrm{in}}$ and final unsqueezing $S(-r_{\mathrm{pre}})$ operations, implemented via $H_{2}$ in the LD regime:
\begin{equation}
    {C'}_{N}^{(3)}(t) = S(-r_{\mathrm{pre}})\,C_{N}^{(3)}(t)\,S(r_{\mathrm{pre}}) \label{eq:presqueezing}.
\end{equation}
By increasing the state's spatial extent in the relevant quadrature, this ``pre-squeezing'' amplifies the accumulation of nonlinear phase terms, making the Hamiltonians more effective and yielding a higher-fidelity cubic gate \cite{GottesmanPRL2001, burd2024squeezing}.
}

\begin{figure}[htbp]
    \centering
    \includegraphics[width=\columnwidth]{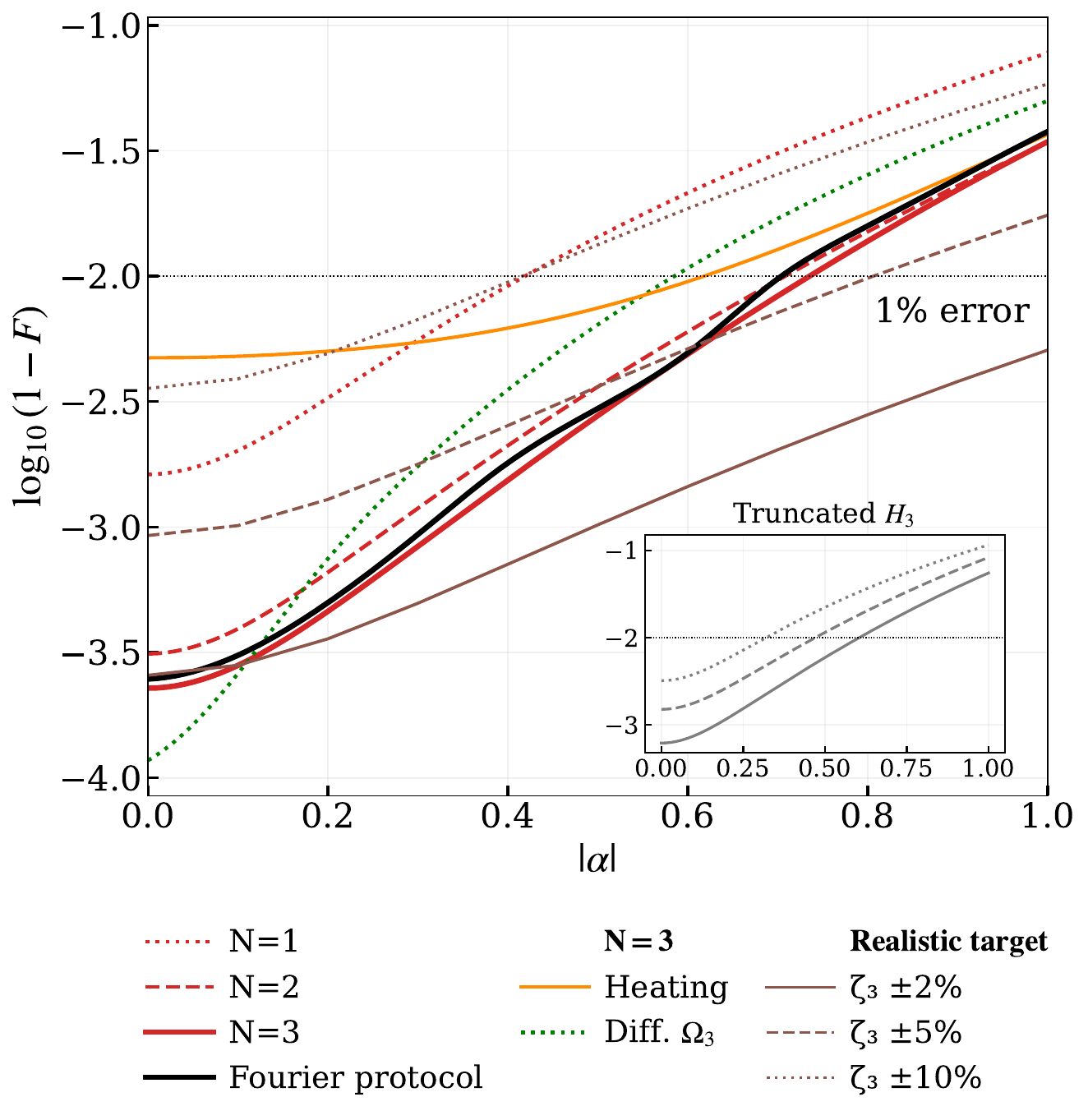}
    \caption{Logarithmic infidelity \(\log_{10}(1 - F)\) as a function of input {coherent state} amplitude \(|\alpha|\) for different gate configurations:  Hamiltonian \(H_{3}\) {in LD limit} (inset), full \(H_{3}\) beyond the Lamb-Dicke limit (red),  and full \(H_{3}\) with heating ($\dot{n}_{\mathrm{th}} = 10 \, \text{quanta/s}$, orange) \cite{Talukdar2016}. Increasing the number of gates systematically improves the fidelity, with the largest gain observed between the truncated and full Hamiltonian cases. The best performance is obtained for $N=3$ {(9 gates with 3 additional displacement)} with an initial {vacuum} state, reaching \(F \approx 0.9998\). 
    The Fourier protocol  curve represents a competing method that synthesizes the cubic interaction from a sequence of $24$ linear operations {(with 24 additional displacement)} \cite{Park2024NPJQIEfficientCubic}.
    For  coherent states with various amplitude $\alpha$, the fidelities by our protocol with $N=3$ and the Fourier protocol  have comparable values.
    The green curve ({$N=3$}) shows a sequence re-optimized specifically for the input {ground} state (\(|\alpha| = 0\)). 
    The dashed lines represent the infidelity of an \textit{ideal} cubic phase gate with $\pm 2\%$, $\pm 5\%$,  and $\pm 10\%$ fluctuations in $\zeta_3$, serving as a baseline to evaluate the protocol's sensitivity against standard calibration errors.
    }
    \label{fig:repetition-fidelity}
\end{figure}

 \textit{{Fidelity of cubic mechanical gate}}
---In this work, we optimize the protocol parameters {$({t'}_1^{(l)}, t_3^{(l)}, t_2^{(l)}, \beta^{(l)})$} in the circuits in Fig. \ref{fig:circuit} and in Eq.(\ref{eq:protocol}) {to synthesize the target gate $U^{(3)}$ by maximizing} 
the fidelity  {of the operation $C^{(3)}_N$ over a range of motional states}
, while keeping the Lamb-Dicke parameter fixed at $\eta = 0.3$, and the Rabi frequencies at $\Omega_{1} = 0.3$ and $\Omega_{3} = 0.3$ (in units of $ \text{MHz}$, unless otherwise specified). 
The Uhlmann-Jozsa fidelity between the generated state $\rho_{\text{gen}}$ and the target state $\rho_{\text{target}}$ $F(\rho_{\text{gen}}, \rho_{\text{target}}) = \left( \mathrm{Tr} \left[ \sqrt{ \sqrt{\rho_{\text{target}} } \, \rho_{\text{gen}} \, \sqrt{ \rho_{\text{target}} } } \right] \right)^2$, 
is optimized via a differential evolution algorithm {(see Appendix \ref{appendix:optimization}.1 for optimization details and \ref{appendix:optimization}.2 for noise modeling)}.

The performance of the {composite gate architecture} as in (\ref{eq:presqueezing}) 
can be quantified by comparing it with the standard single-gate protocol {$C_{N=1}^{(3)}$}. 
With  optimized {pre}-squeezing of 6.71\,dB ($r_{\text{in}} \approx 0.772$) {incorporated as a gate sub-routine,} 
the {gate} fidelity of the {cubic gate} {benchmarked on motional ground state} improves from $F = 0.998$ to  $F = 0.9996$,  corresponding to the 5-fold reduction of infidelity for $N=1$. 
While this composite pre-squeezing sequence can enhance fidelity and shortens pulse duration for low-depth ($N=1$) sequences, gate fidelity performance saturates and timing benefits diminish for $N \geq 2$. Thus, for simplicity, our primary benchmark utilizes the straightforward extension of $C_{N=3}^{(3)}$ as defined in Eq. (\ref{eq:protocol}).
{In Table \ref{tab:params_3g} are summarized all optimized parameters by our protocol for $N=3$. We note that there is an even solutions with higher fidelities at a larger total gate duration.  By further expanding the optimization parameter space, we achieve even higher fidelities of $F = 0.999975$ for the vacuum state and $F = 0.9992$ for $\alpha = i$, at the cost of twice  longer total gate duration of $1.5$ ms. This illustrates a tunable trade-off between gate speed and precision within our protocol.}
 
{\textit{Mechanical Wigner function.}}---Fig.~\ref{fig:wignercomp} compares the Wigner functions of the ideal target state and the optimized $N=3$ generated state.
The generated state in Fig.~\ref{fig:wignercomp}(b) achieves an exceptionally high fidelity of $F=0.99986$ with the target in Fig.~\ref{fig:wignercomp}(a) {on vacuum state input}. Crucially, our protocol accurately reproduces the complex, non-Gaussian structure of {this} cubic phase state, including its characteristic parabolic crescent shape and {multiple} prominent regions of negativity---a key signature of {advanced quantum non-Gaussianity from nonlinear cubic phase gate}. 

Figure \ref{fig:wignercoherent} displays the Wigner functions for inputs with $|\alpha|=1$ along different axes ($\alpha \in \{\pm 1, \pm i\}$). The generated states faithfully reproduce the {output states from} target cubic phase gate's {inducing a position-dependent momentum shift that curves the Gaussian profile into the characteristic crescent shape.
For real displacements $\alpha = \pm 1$, the fidelity remains near-unity ($F > 0.999$), whereas for imaginary displacements $\alpha = \pm i$, the fidelity decreases to $F \approx 0.969$. This reduction indicates that states displaced in momentum are more sensitive to residual non-commuting terms in the full Hamiltonian (Eq. (\ref{eq:H1}-\ref{eq:H3})) that are not fully compensated by the optimized sequence.
}

The Wigner functions provide the {comprehensive} evidence of success of our protocol, {qualitatively} confirming the faithful synthesis non-Gaussian interference fringes {caused by} crescent phase-space structures.

{Figure \ref{fig:density_wigner} quantifies the gate's convergence by comparing Wigner negativity cuts along the position axis $p=0$. As the repetition number $N$ increases, the generated state's negativity depth and oscillation frequency progressively match the target. Figure \ref{fig:diagcut} further validates the protocol in the worst-case scenario ($\alpha = -i$), where the state is displaced along the axis of maximal distortion. The cut along the diagonal of maximal negativity reveals that the $N=3$ protocol  qualitatively reproduces the high-frequency interference fringes and the specific negativity minima ($x \approx -10$), confirming  phase-matching between the $H_{1}$ and $H_{3}$ interactions.}

Figure~\ref{fig:repetition-fidelity} compares the performance of the gate architecture {in Fig. \ref{fig:circuit}} under two regimes. 
In the inset panel, we restrict the third-sideband Hamiltonian $H_{3}$ to the Lamb-Dicke regime by truncating higher-order terms {$\mathcal{O}(\eta^5)$}. 
Under this approximation, $N=1$ protocol yields limited fidelity at larger input coherent state amplitude $|\alpha|\approx 1$, but repeating the gate with independently optimized parameters allows the nonlinearity to accumulate effectively, reaching $F \approx 0.9992$ after three {repetitions} {for input ground  state}.
{It yields $F=0.884$ for $|\alpha| \approx 1$, while $N=3$ improves this to $F=0.9447$, showing a much faster enhancement than for vacuum.}
The low fidelity at a large $\alpha$ is attributed to the cubicity change. {We identify this shift by finding the effective cubicity $\zeta_3^{eff} \equiv \xi_{min}$ that minimizes the nonlinear variance to be discussed later. Crucially, we find that $\zeta_3^{eff}$ scales monotonically and slightly non-linearly with the input amplitude $|\alpha|$. 
{The shift scales as $\zeta_3^{eff} \approx \zeta_3(1 + \gamma \eta^2 |\alpha|^2)$, where $\gamma$ represents the fifth-order contribution that can be numerically found, which was derived in Appendix \ref{append:effcub}.}
When the generated state is compared to an ideal gate with this optimized strength $\zeta_3^{eff}$, the fidelity remains high for coherent states with $|\alpha|=1$ ($\approx 0.992$), confirming that the error is a more predictable parameter shift rather than a loss of non-Gaussian structure.} 
In the main panel, we show results obtained using the full Hamiltonian beyond the Lamb-Dicke limit, where {all} higher-order terms in $H_{3}$ are retained. 
While the qualitative behavior remains similar, the full model consistently achieves higher fidelities across the entire $|\alpha|$ range, with fidelities up to $F \approx 0.9998$ {at input vacuum}   {for $N=3$}.
{For $|\alpha| \approx 1$, the protocol without truncation also achieves a significant enhancement, rising from $F=0.9222$ ($N=1$) to $F=0.9655$ ($N=3$).}
We also highlight the additional curve (green) obtained by increasing the first-sideband Rabi frequency 
and 
re-optimizing the {$N=3$} sequence. 
This specific optimization illustrates the protocol's flexibility: by tailoring the {Rabi frequencies} 
fidelity can be  improved {for small $|\alpha|<0.1$}, albeit at the cost of performance for {larger} coherent state inputs.
{To evaluate the protocol's performance under realistic conditions accounting for typical cubicity fluctuations, we compare the generated gates against ideal ones} exhibiting varying cubicity $\zeta_3$.
We include a region of cubicity fluctuations  representing ideal cubic states with $\zeta_3 \in [0.9, 1.1]$, $[0.95, 1.05]$, and $[0.98, 1.02]$. This comparison reveals that the {realistically achievable} fidelity decrease for larger $|\alpha|$ is linked to the sensitivity of the high-frequency interference fringes observed in Figs. \ref{fig:wignercoherent}-\ref{fig:diagcut}. As the input amplitude increases, the input state probes regions where even minor residual phase-space shifts result in a significant drop in fidelity. 
While the $N=3$ protocol outperforms an ideal gate with 2\% calibration error at $|\alpha|=0$, its infidelity increases more sharply at larger amplitude $|\alpha|$. For $|\alpha| > 0.2$, the protocol becomes more sensitive than the 2\% error baseline. 
\textcolor{black}{This steeper slope than the baseline due to higher-order distortions can be mitigated by cancellation of them \cite{SametiPRA2021}. } 
As established above, residual higher-order nonlinear distortions induce this effective shift in the cubicity strength, leading to the observed mismatch between the generated gate and the fixed target gate $\zeta_3 = 1$. 
The performance of the gate for higher coherent amplitudes ($\alpha=2$) is further analyzed in Appendix \ref{appendix:wignerlarge}.

Interestingly, we found that optimizing the gate parameters using an input thermal state, rather than the vacuum state, significantly enhances the gate's performance across a broader phase space,  i.e. improved fidelities for input coherent states with larger amplitudes $|\alpha|$. 

This approach {can be further used for} benchmarking {nonlinear phase gates on} coherent states with a Gaussian distribution of displacements~\cite{Krauter2013}. 

We also {test} our protocol against the recent  {proposal}~\cite{Park2024NPJQIEfficientCubic}, which realizes the cubic phase gate through a sequence of 24 applications of the first-sideband Hamiltonian $H_{1}$ {(with additional \cmmnt{24} displacement {gates analogical to $U_1$})}. 

In contrast, our $N=3$ protocol {certainly} achieves {comparable} fidelity across the range of  input coherent amplitudes  while requiring only 9 gate applications (considering the first $U_1$ can be {alternatively} applied by a classical laser drive), a near three-fold reduction in the number of sequential gate applications. 
{By utilizing parametric trap modulation for the squeezing $U_2$, the requirement for sideband-driven pulses is further reduced. This laser-free implementation of $U_2$ allows the protocol to rely solely on the first and third sidebands for the non-Gaussian accumulation, increasing the enhancement of number of sideband pulses to 12 folds. In addition, the total time is also reduced from $0.562$ ms, thus narrowing further the gap. }

We note a trade-off between gate duration and control complexity when comparing {the method in Fig. \ref{fig:circuit} and in the Ref.~\cite{Park2024NPJQIEfficientCubic}}. Based on the experimental parameters ($\Omega_0 = 2\pi \times 250\,\text{kHz}$, $\eta=0.3$), the Fourier protocol achieves the target interaction strength in a total interaction time of $T_{\text{Fourier}} \approx 72\,\mu\text{s}$, if we choose the same $\eta=0.3$. In contrast, our direct synthesis method requires a total gate time of $T_{\text{total}} \approx 797 \,\mu\text{s}$. However, the Fourier method traditionally requires a much smaller $\eta$ to remain within the Lamb-Dicke limit, which would significantly increase its total interaction time \cite{SametiPRA2021}. 
{Our numerical analysis reveals that the Fourier infidelity grows as $\mathcal{O}(\mathcal{\eta}^4)$ and $\mathcal{O}(\mathcal{N}^4)$ where $\mathcal{N}$  is the number of gates due to the coherent accumulation of non-commuting higher-order residuals. To achieve comparable fidelity, the Fourier protocol would require a drastically smaller $\eta$ and suppressed Rabi frequencies, leading to a longer total interaction times.}
{Infidelity of each gate  $1-F\approx 10^{-4}$ requires a low LD parameter $\eta\approx 0.07$}, that increases the total time close to $T_{\text{Fourier}} \approx 300\,\mu\text{s}$.  Furthermore, as such an error accumulate rapidly,  the actual $\eta$ may need to be much lower, i.e. the total time even longer. 
While the Fourier protocol \cite{Park2024NPJQIEfficientCubic} requires calibrating a sequence of 24 (or 48 if displacement is supplied by first resonant sideband pulse) distinct linear operations (each  requiring pulse area and phase calibration), our direct synthesis method achieves the same nonlinearity with only 9 (or 12 if displacement is supplied by first resonant sideband pulse, {or 6 if parametric squeezing is exploited}) composite pulses. 
Our protocol thus reduces calibration complexity,  trading  {competitive} interaction times {at high fidelity} for a significant reduction in control-pulse overhead. 

{\textit{Robustness of mechanical gate}.---}
To assess the protocol's robustness against imperfections, we model the effects of motional heating ($\dot{n}_{\mathrm{th}} = 10$ quanta/s) and motional dephasing ($T_{\mathrm{coh}} = 50$ ms) \cite{Talukdar2016}. As shown in Fig.~\ref{fig:repetition-fidelity}, infidelity increases with gate depth due to cumulative noise exposure;however, the Wigner functions remain qualitatively consistent with the target, preserving the non-Gaussian interference structure and negative regions (Appendix \ref{appendix:disspation}). For the heating model, the protocol achieves a fidelity of $0.993$ for an initial vacuum state ($\alpha = 0$) and $0.962$ for $\alpha = i$. Under motional dephasing, the fidelity for the vacuum state is approximately $0.971$, falling to $0.89$ for larger amplitudes such as $\alpha = \pm i$, having a much stronger impact than the heating. Despite these factors; here as well, the generated states preserve their characteristic non-Gaussian structure and phase-space negativities. Finally, we evaluated static calibration errors by introducing independent $\pm 1\%$ perturbations to all gate parameters (Table~\ref{tab:params_3g}).
This resulted in a negligible decrease in fidelity (from $F \approx 0.9998$ to $F \approx 0.9991$), demonstrating the protocol's resilience to minor control field imperfections.

{To quantify the necessity of the third sideband interaction, 
we benchmarked our protocol against a simplified variant} that excludes the third-sideband unitary $U_{3}$ from the circuit in Fig.~\ref{fig:circuit}, relying solely on $H_{1}$ and $H_{2}$ {beyond LD regime} \cite{McDonnell2006PhysRevLett.98.063603LambDicke}. 

{Our numerical analysis reveals that while this approach suffices for weak nonlinearities, it degrades significantly as cubicity increases {due to the inefficient scaling of the parasitic $i(\hat{a}^2\hat{a}^\dagger - \hat{a}^{\dagger 2}\hat{a})$ term within $H_1$.} 
This justifies the direct exploitation of $H_3$ as the primary driver for high-fidelity, strong nonlinear gates.}

\textit{Nonlinear {mechanical} squeezing} --- Nonlinear squeezing \cite{Moore2022QNG,KalaOptExp2022Nonlinearsqueezing} occurs when quantum noise in {the nonlinear function of quadrature operators} is reduced through higher-order Hamiltonians \cite{RakhubovskyFilip2021Stroboscopic,MooreNJP2019Estimation}, such as {for the cubic gate} (\( H \approx \hat{X}^3 \)) or {the quartic gate} (\( H \approx \hat{X}^4 \)). 
As a crucial feature of non-Gaussian quantum states, nonlinear variance quantifies the reduction of fluctuations for nonlinear phase operators  like $\hat{P} + j \xi\hat{X}^{j-1}$  for a real number $\xi$. {As was noted before,} the value at which  this variance is minimized defines the effective nonlinearity strength $\zeta_j^{\text{eff}} \equiv \xi_\mathrm{min}$.

\begin{figure}[h]
    \centering
    \includegraphics[width=\columnwidth]{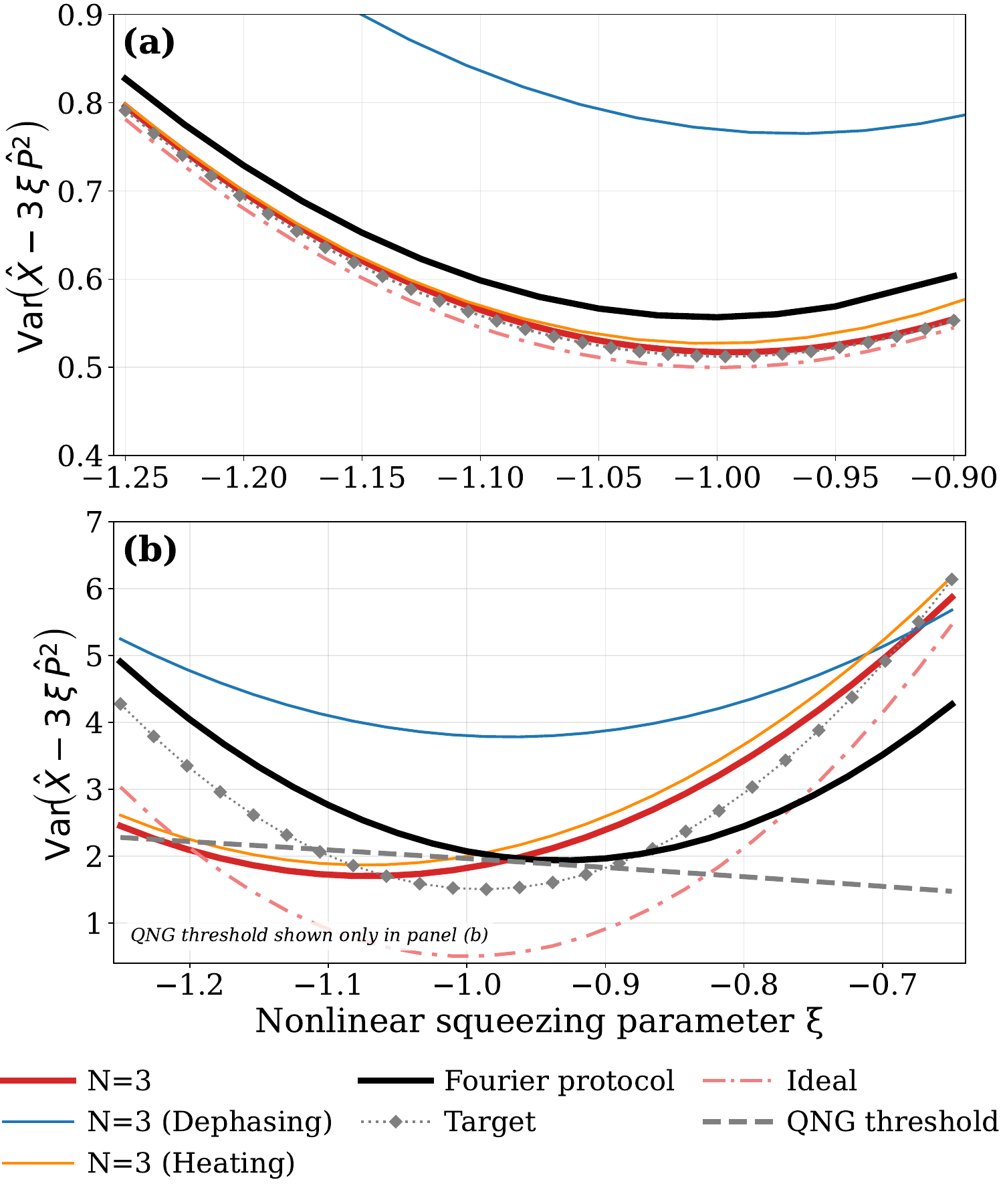}
    \caption{
        Nonlinear variance $\mathrm{Var}(\hat{X}-3\xi\hat{P}^2)$
        for cubic gate acting on (a): the {ground state}, and
        (b): a coherent state with imaginary amplitude $\alpha=i$.
        The {analytical cubic gate (pink dot-dashed)} represents the theoretical bound,
        while the {numerical target (grey dotted)} shows the effect of finite-basis truncation.
        The {optimized $N=3$ protocol (red solid)} achieves a minimum below the
        {QNG threshold ({$V_{QNG} \approx 1.965$ at $\xi=1$}, grey dashed)}, whereas the {Fourier-synthesis benchmark (black dotted)} remains above the threshold
        for the coherent state.
        {Resilience is shown for motional dephasing (blue) and motional heating (orange)}.
    }
    \label{fig:nonlinear_variance_cubic_gate}
\end{figure}

\begin{figure}[t!]
    \centering
    \includegraphics[width=\columnwidth]{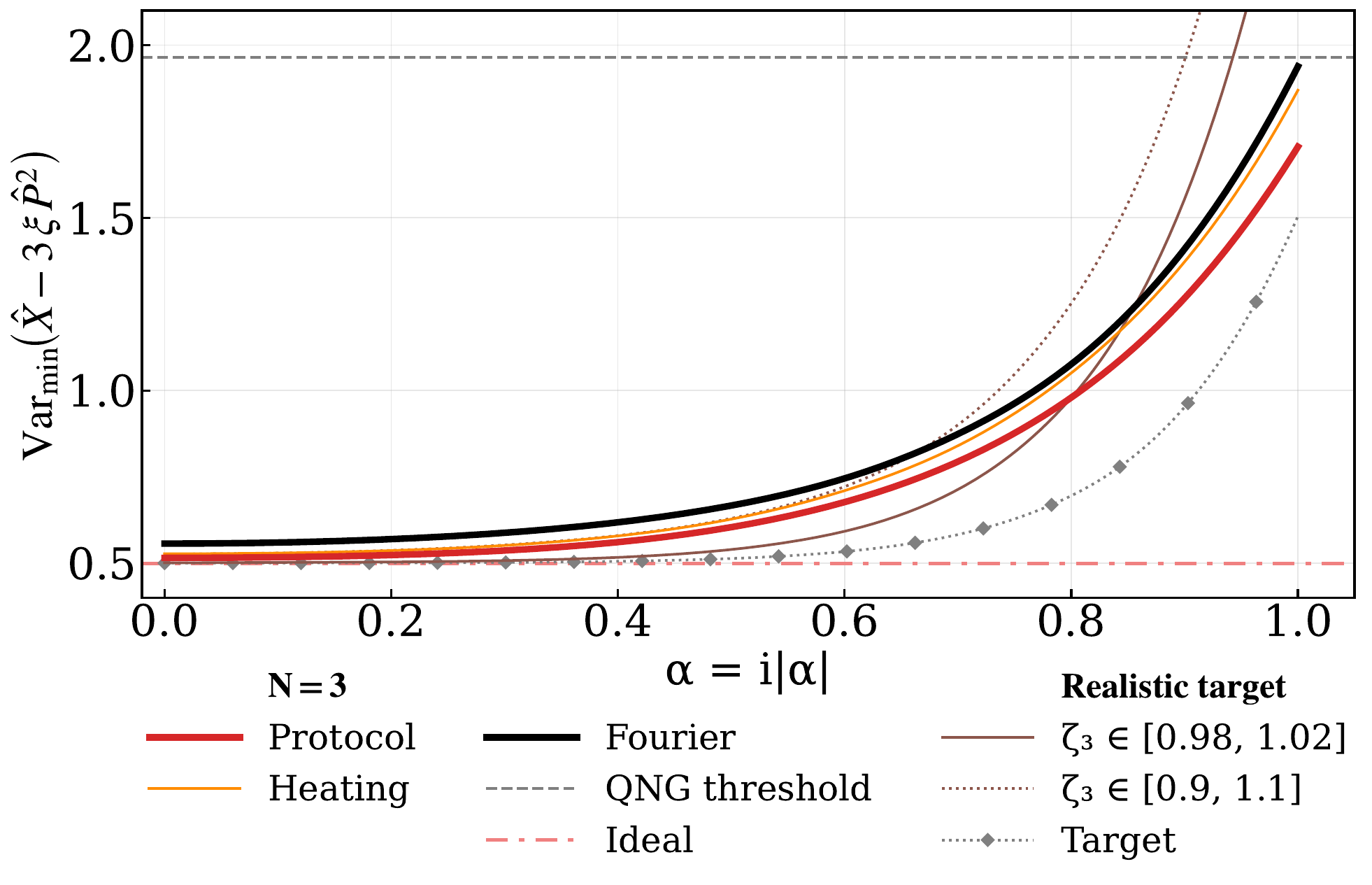}
    \caption{Comparison of the minimum nonlinear variance $\mathrm{Var}(X - 3\xi P^2)$ {optimized over $\xi$} for the cubic phase gate acting on coherent states with imaginary amplitude $\alpha$ {without loss and noise}. Our optimized $N=3$ protocol (red) yields a {slightly} lower variance than the Fourier-method benchmark (black) over the entire range of amplitudes and approaches the theoretical limit (pink dot-dashed) for small $\alpha$. {The realized cubicity $\zeta_3^{\text{eff}} \equiv \xi_\mathrm{min}$ is close to target value $\zeta_3=1$, e.g. $\approx 1.015$ at $\alpha=0.5i$, while $\approx 1.08$ at $\alpha=i$.} The quantum non-Gaussian (QNG) threshold (gray dotted) is also shown. At $\alpha = 1$, the variance obtained with our protocol lies approximately $13\%$ below the QNG threshold \cite{Moore2022QNG}, certifying non-Gaussianity. Extrapolating the trend, the crossing point with the QNG threshold occurs at $\alpha \simeq 1.045$.
    {The gray dotted line indicates the QNG threshold ($V_{QNG} \approx 1.965$). The brown  lines denote the nonlinear variance of ideal states with $\pm 10\%$ (dashed) and $\pm 2\%$ (solid) cubicity fluctuations, showing that the minimum nonlinear variance increases with $\alpha$.}
}
    \label{fig:nonlinear_squeezing}
\end{figure}

Figure~\ref{fig:nonlinear_variance_cubic_gate} compares the nonlinear variance $\mathrm{Var}(\hat{X}-3\xi\hat{P}^2)$ {to align with the implementation of the cubic gate $e^{i\zeta_3\hat{P}^3}$} for our protocol, the Fourier protocol, and the ideal cubic phase gate, revealing that our $N=3$ sequence  faithfully reproduces the target characteristics for both {ground state} and coherent state ($\alpha=i$) inputs. For {ground state in Figure~\ref{fig:nonlinear_variance_cubic_gate} (a)}, the generated variance is nearly indistinguishable from the ideal, with minima differing by less than $0.03$. 
{In Fig. \ref{fig:nonlinear_variance_cubic_gate}(b), the $N=3$ protocol achieves a minimum variance of $1.706$ at $\xi \approx -1.082$,  outperforming the Fourier benchmark which only reaches the minimum variance of $\approx2.20$. While the target gate reaches a lower minimum of 1.505 at $\xi \approx -0.986$, our protocol maintains the non-Gaussian signature below the QNG threshold ($V_{QNG} \approx 1.965$), which the Fourier protocol fails to reach. The shift in the optimal $\xi$ compared to the ideal case suggests that residual higher-order interactions introduce a modified effective cubicity. {Crucially, the minimum variance of our $N=3$ gate matches the numerical target gate precisely, implying its high quality except for the numerical limitations.} 
The discrepancy between the analytical ideal and the numerical target stems from the finite Hilbert space dimension used in the simulations.  In experimental realizations, the achievable nonlinear squeezing may be lower than these numerical benchmarks due to physical constraints.}

To extend this analysis beyond specific instances {of $\alpha$}, Fig.~\ref{fig:nonlinear_squeezing} illustrates the minimum nonlinear variance $\min_{\xi} \mathrm{Var}(\hat{X} - 3\xi \hat{P}^2)$  as a function of the input coherent state amplitude {$\alpha=i|\alpha|$}. The results confirm that our optimized $N=3$ protocol without dephasing and heating (blue curve)  yields a {slightly} lower variance than the Fourier-method benchmark (red curve) across the entire range of amplitudes. 
{We} evaluate the generated states against the hierarchy of {nonlinear squeezing for} conservative motion  \cite{Moore2022QNG} {(it coincides with \cite{KalaOptExp2022Nonlinearsqueezing})}. 
We identify two distinct thresholds: the classical threshold for coherent states, $V_{\text{cl}}$, and the quantum non-Gaussian threshold, $T_{\text{QNG}}$.
For a cubic interaction corresponding to $\xi \approx 1$, the classical {nonlinear variance} scales quadratically as $V_{\text{cl}}(\alpha) \approx 5.0 + 36\text{Im}(\alpha)^2$ {for coherent states}, {while nearly constant over real part of $\alpha$}. In contrast, the QNG threshold, which represents the minimum variance achievable by \textit{any} Gaussian state, is significantly lower at $T_{\text{QNG}} \approx 1.965$ {at $\xi=1$} \cite{Moore2022QNG, KalaOptExp2022Nonlinearsqueezing} and is invariant under displacement $\alpha$. 
{To compare with a realistic target, we consider the} minimum nonlinear variance for ideal states with fluctuation in cubicity that exhibits a similar rise in nonlinear variance with $\alpha$ when calculated numerically. 
Generally for other values of $\xi$'s, it is given by a formula $T_\mathrm{QNG}=\frac{3}{2^{5/3}}(3\xi)^{2/3} $ \cite{Moore2022QNG,KalaOptExp2022Nonlinearsqueezing}.
While sensitive to higher-order sideband distortions at large $\alpha$, the nonlinear variance remains below the QNG threshold, validating the cubic character more accurately than fidelity. 
Differently than for the fidelity in Fig. \ref{fig:repetition-fidelity}, the protocol shows a better slope in non-linear squeezing for larger $|\alpha|$ than  ideal gates with varied cubicity.
This resilience is confirmed by the preserved interference structures by the preserved interference structures in Figs.~\ref{fig:wignercomp} and \ref{fig:wignercoherent},  even for states with higher motional energy.
This decrease in fidelity for imaginary amplitudes $\alpha$ indicates that average quantities like nonlinear variance and fidelity are sensitive to  {the initial motional excitation of the input state. 
}

\begin{figure}[htbp!]
    \centering
    \includegraphics[width=\columnwidth]{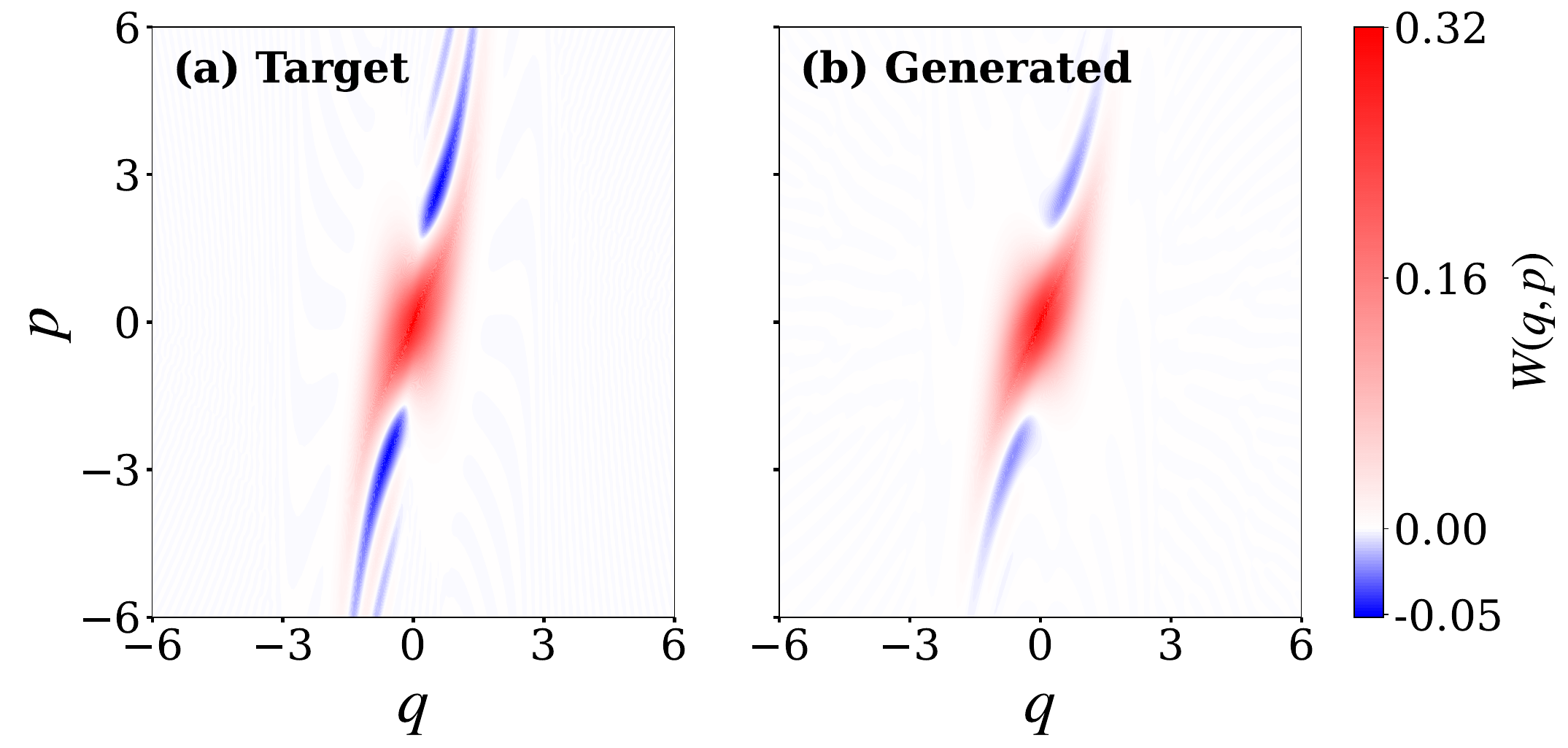}
    \caption{{Wigner function representations of (a) the ideal quartic-phase state with quarticity $\zeta_4 = 0.25$ prepared from the vacuum input, and (b) the state generated using the asymmetric-squeezing sequence
    $S(r_{\mathrm{post}}),R(\theta),U_4(T_4,\phi_4),U_2(T_2,\phi_2),S(r_{\mathrm{pre}})$
    with $\Omega_2 = 0.2$ and $\Omega_4 = 0.8$.
    The generated state exhibits the characteristic four-fold non-Gaussian interference pattern of the quartic gate, closely matching the target with a fidelity of $F = 0.995$.}}
    \label{fig:quartic}
\end{figure}

To further demonstrate {how} our approach generalizes beyond the cubic case,
we implemented an extended sequence for realizing a quartic-phase gate $\mathcal{U}^{(4)}=e^{i \zeta_4 \hat{X}^4}$,
\begin{align}
C^{(4)}(t)
&= S(r_{\text{post}})R(\theta)U_4(t_4,\phi_4) \nonumber\\
&\quad\times U_2(t_2,\phi_2)S(r_{\text{pre}}).
\end{align}
acting on the motional vacuum.
{Here, all parameters $t_k$, $\phi_k$, $\theta$, $r_\mathrm{pre}$ and $r_\mathrm{post}$ are optimized.}
Here, $U_2$ and $U_4$ correspond to two-tone drives  pulses resonant with the second and fourth sidebands, with respective Rabi frequencies $\Omega_2 = 0.2$ and $\Omega_4 = 0.8$, while $S(r_{\mathrm{pre}})$ and $S(r_{\mathrm{post}})$ are pre- and post-squeezing operations that compensate residual phase-space distortions, and $R(\theta)$ denotes a rotation in phase space.
This protocol successfully generates the characteristic four-fold interference structure of the quartic-phase state with quarticity $\zeta_4 = 0.25$, achieving a fidelity of $F = 0.995$.
This \textcolor{black}{qualitative capture of core quartic features} is visually confirmed in Fig.~\ref{fig:quartic}, which compares the Wigner function of the ideal target state with the one produced by our protocol. 
The generated gate in Fig.~\ref{fig:quartic}(b) captures the central non-Gaussian structure and the first negative regions for a single-round ($N=1$) protocol, confirmed by the high fidelity ($F=0.995$). However, resolving the second negative area of ideal quartic gate in Fig.~\ref{fig:quartic}(a)—essential for the full interference character of a quartic resource—remains numerically challenging at this depth.  Extending the protocol to higher $N$ is required to fully resolve these high-frequency structures.

\textit{Discussion}---
While the global metrics such as  fidelity in Fig. \ref{fig:repetition-fidelity} and nonlinear squeezing in Fig. \ref{fig:nonlinear_squeezing} are quantitatively distinct from the {ones for the ideal gate} because residual nonlinearities shift the effective cubicity, the protocol robustly maintains the qualitative Wigner negativity and interference fringes in Fig. \ref{fig:wignercomp} and \ref{fig:wignercoherent}, demonstrating that our protocol effectively reaches the non-Gaussian regime required for utility. 
{These essential phase-space effects can enable the experimental investigation of compound nonlinear processes that remain unexplored \cite{Moore2025Nonlinear}.}
The systematic convergence of the density matrix elements toward the target state is documented in Appendix \ref{appendix:convergence}.

{
Our efficiency stems from directly exploiting the higher-order nonlinearities in Eqs.~(\ref{eq:H1}-\ref{eq:H3}) that are  present beyond the Lamb-Dicke regime.  Unlike previous methods \cite{Park2024NPJQIEfficientCubic} which utilized only the resonant first-order displacement term $\propto \eta$ in $H_{1}$ while suppressing all higher-order contributions, we harness the native $\mathcal{O}(\eta^3)$ cubic term in Eq.~(\ref{eq:H1}) and the leading $\mathcal{O}(\eta^3)$ term in $H_{3}$ [Eq.~(\ref{eq:H3})] as primary gate primitives.} This offers a structured control toolbox for non-Gaussian operations, although its scalability is bounded by physical constraints. As the order $j$ increases, the effective coupling strength scales as $\mathcal{O}(\eta^j)$, necessitating either longer gate durations or increased laser intensities. In practice, higher intensities can worsen off-resonant carrier coupling and AC Stark shifts, while larger $\eta$ increases sensitivity to spectator motional modes. ;

Experimental feasibility {with trapped ions} relies on managing AC Stark shifts and motional heating. Although driving higher-order sidebands requires stronger gradients ($\Omega_{3} \approx 0.3$ at $\eta=0.3$), standard auxiliary tone or tensor-shift cancellation techniques can compensate for Stark shifts \cite{Gaebler2016}. Furthermore, assumption of negligible thermalization is supported by recent cryogenic trap benchmarks ($< 1$ quanta/s heating rates) \cite{Brownnutt2015}. Crucially, resonant bichromatic scheme creates a closed subsystem, avoiding the multiple adiabatic ramps required by older sequential monochromatic schemes \cite{Rangan}.
Our method offers a hardware-efficient alternative to Fourier-like synthesis \cite{Park2024NPJQIEfficientCubic}. By using higher-order terms as primitives rather than errors, we achieve a three-fold reduction in sequential sideband pulses and \textcolor{black}{mitigate} the error scaling characteristic of linear-decomposition methods at high $\eta$. This structural stability allows for deterministic operation in the non-Lamb-Dicke regime where previous protocols diverge. 
On the other hand, the Fourier synthesis protocol requires a much smaller Lamb-Dicke parameter to avoid the accumulating error. Similarly, while other trapped-ion schemes engineer effective nonlinearities via Magnus expansions of linear forces \cite{Bazavan2024squeezing}, we utilize the inherent nonlinearity of the ion-light interaction itself. Unlike state conversion protocols \cite{zheng2021}, we synthesize the gate from fundamental dynamics without requiring pre-existing non-Gaussian resources. Finally, while analogous native nonlinearities exist in superconducting SNAIL elements \cite{Eriksson2024universal}, our source of nonlinearity arises from intrinsic light-matter physics rather than engineered circuit potentials.

\textit{Conclusion---}
We demonstrated an efficient, deterministic protocol exploiting sideband interactions beyond Lamb-Dicke regime for synthesizing high-fidelity cubic phase gates {that accurately reproduce essential non-Gaussian interference features in the phase space.}

{By quantitative analysis we verify that this} methodology establishes a robust, hardware-efficient framework for universal continuous-variable quantum processing in trapped-ion systems \cite{LloydPRL1999,GottesmanPRL2001}.

{Future implementations should leverage comprehensive numerical optimization of multi-sideband control fields derived from the general Hamiltonian. Experimentally, the focus must be on maintaining precise control of two-tone drives and mitigating motional decoherence for larger quantum states.}

\section*{Code and Data Availability}
The source code used for the numerical simulations and figure generation is available at \url{https://github.com/akramkasri/CUBIC_Gate_optimiser}.

\section*{Acknowledgment}

A.K., K.P. and R.F. received support by project CZ.02.01.01/00/22\_008/0004649 (QUEENTEC) of EU and MEYS Czech Republic.
K.P. and R.F. acknowledge the grant no. 22-27431S of Czech Science Foundation. R.F. was also supported by the project No. LUC25006 of MEYS Czech Republic.
We thank Kratveer Singh, Lukáš Slodička, and Darren W. Moore for helpful discussions.

\section*{Author contribution}

R.F. supervised the project. A.K., K.P., and R.F conceived the theoretical idea.
A.K. performed the simulations with the inputs and feedbacks from K.P. and R.F. 
K.P. and A.K. derived the analytical models with the inputs and feedbacks from R.F. 
All authors contributed to the manuscript preparation and discussed the results.

\bibliographystyle{unsrt}
\bibliography{references}

\onecolumn\newpage
\appendix

\section{Numerical Optimization and Noise Modelling}
\label{appendix:optimization}

\subsection{Optimization Algorithm}

The control parameters for each gate block, $\{t_{1}, t_{3}, r, \beta\}$, are optimized using a differential evolution (DE) algorithm. The objective function to be minimized is the negative fidelity between the state generated by the protocol, $\rho_{\mathrm{gen}}$, and the target cubic-phase state, $\rho_{\mathrm{tgt}}$:
\[
\mathcal{L}(\mathbf{p}) = -F(\rho_{\mathrm{gen}}(\mathbf{p}), \rho_{\mathrm{tgt}}),
\]
where the fidelity is evaluated numerically using \textsc{QuTiP}. The state is constructed sequentially as $\psi \mapsto G_{N} \cdots G_{2} G_{1} \psi$, where each gate block is defined as $G_{k} = S(r_{k}) U_{1}(t_{1,k}) U_{3}(t_{3,k}) D(\beta_{k})$.

The DE search is conducted within the bounded parameter domain:
\[
t_{1}, t_{3} \in [0, 300], \quad r \in [-2, 2], \quad \beta \in [-5, 2],
\]
using a population size of 40, a mutation factor between $0.5$ and $1.0$, and a crossover probability of $0.7$. The optimization is terminated when the change in the best fidelity falls below a threshold of $10^{-6}$ for several consecutive iterations.

\subsection{Noise Modelling}
{To accurately model the experimental imperfections relevant to a trap setup, we incorporate two dominant decoherence mechanisms: motional heating and motional dephasing. The numerical evolution is performed using the \texttt{mesolve} library in \textsc{QuTiP}, which integrates the master equation over the gate duration.}

\paragraph{Motional Heating}
We model the heating of the mechanical mode as a stochastic process resulting from electric field noise on the trap electrodes. This is described by the heating rate $\dot{n}_{th} = 10$ quanta/s \cite{Talukdar2016}. In our simulations, we implement this via a time-dependent master equation where the transition rate into higher Fock states is constant.

To maintain consistency with our gate-block decomposition, we utilize a Kraus representation for each time step $\Delta t$ within the Suzuki--Trotter expansion:
\begin{align*}
K_{0} &= \sqrt{1 - \dot{n}_{th} \Delta t} \, \mathbb{I}, \\
K_{1} &= \sqrt{\dot{n}_{th} \Delta t} \, a^{\dagger}.
\end{align*}
This model captures the diffusive growth of the phonon number during the gate, which induces a random walk of the motional state in phase space.

\paragraph{Motional Dephasing}
Motional dephasing, primarily caused by fluctuations in the trapping potential and RF instabilities, leads to a loss of phase coherence in the harmonic oscillator. We quantify this using a characteristic coherence time $T_{coh} = 50$ ms, which corresponds to a dephasing rate $\gamma_{\phi} = 1/T_{coh} = 20$ s$^{-1}$.

This decoherence is implemented by applying a phase-damping channel at the end of each gate operation. The effect is represented by the Kraus map:
\[
\mathcal{E}_{deph}(\rho) = \sum_{k} E_k \rho E_k^\dagger,
\]
where the operators $E_k$ enforce a decay of the off-diagonal elements of the density matrix in the Fock basis. Specifically, the coherence between states $|n\rangle$ and $|m\rangle$ decays at a rate proportional to $(n-m)^2$. This mechanism is responsible for the "blurring" and smoothing of the Wigner function negativities observed in our phase-space analysis.

The combination of these heating and dephasing models provides a realistic assessment of the protocol's robustness, demonstrating that the multi-sideband gate maintains high fidelity under standard laboratory noise conditions.  

\section{Wigner Function with a larger $\alpha$}
\label{appendix:wignerlarge}

\begin{figure}[h]
    \centering
    \includegraphics[width=\columnwidth]{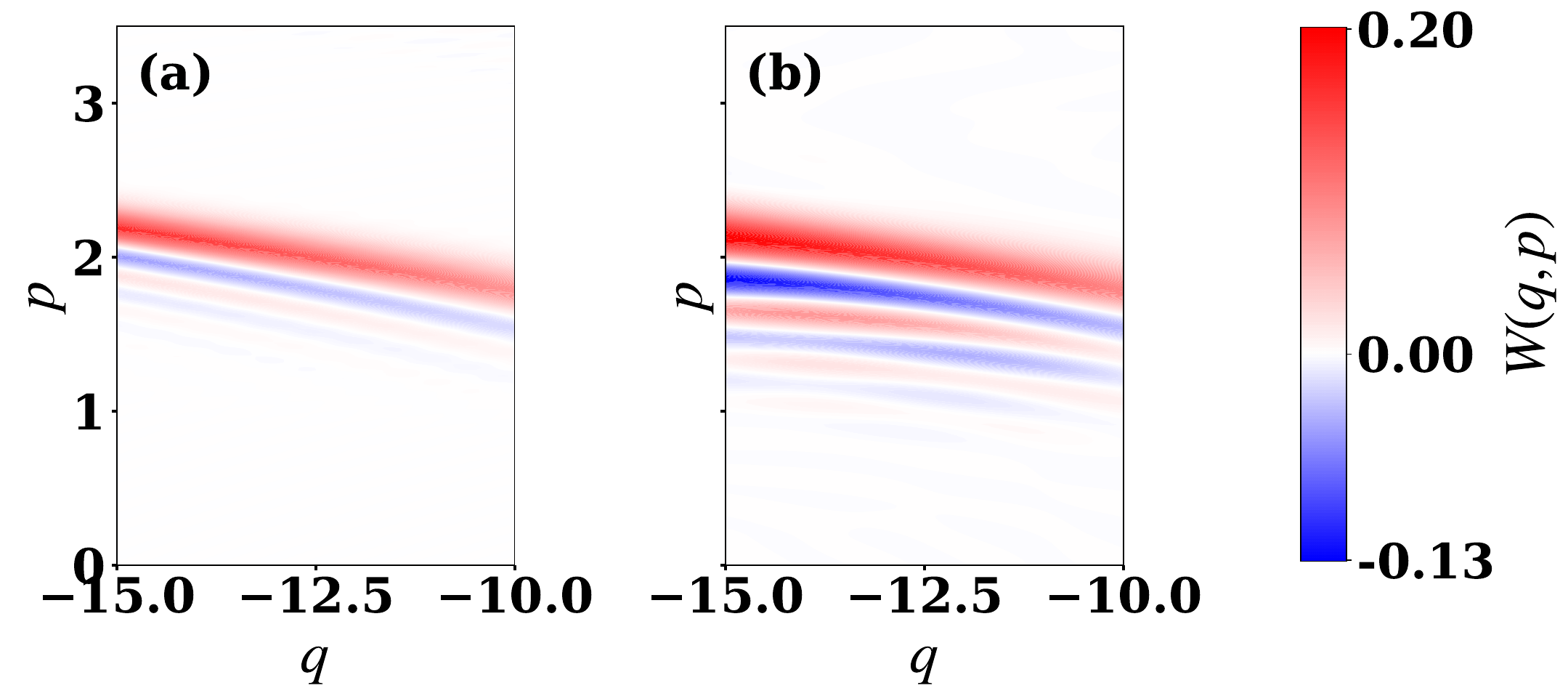}
    \caption{
    Wigner function cuts for a coherent input state with amplitude $\alpha = 2$ after application of the cubic phase gate.
    (a) Target state obtained from the ideal cubic phase operator.
    (b) State generated by the optimized \(N=3\) protocol.
    Compared to the target, the generated state exhibits additional interference fringes and enhanced regions of Wigner-function negativity, reflecting the contribution of higher-order nonlinear terms captured by the protocol.
}

    \label{fig:wigner_nontruncated}
\end{figure}

Figure~\ref{fig:wigner_nontruncated} compares the action of the cubic phase gate on a coherent input state with $\alpha = 2$.
While the target state already displays pronounced non-Gaussian features, the generated state shows additional interference fringes and stronger Wigner-function negativities.
These differences arise from higher-order nonlinear contributions present in the implemented dynamics, which become more pronounced at larger input amplitudes.

\section{Phase-Space Representation under Dissipation and Calibration fluctuation}
\label{appendix:disspation}

\begin{widetext}

\begin{center}
\includegraphics[width=0.95\textwidth]{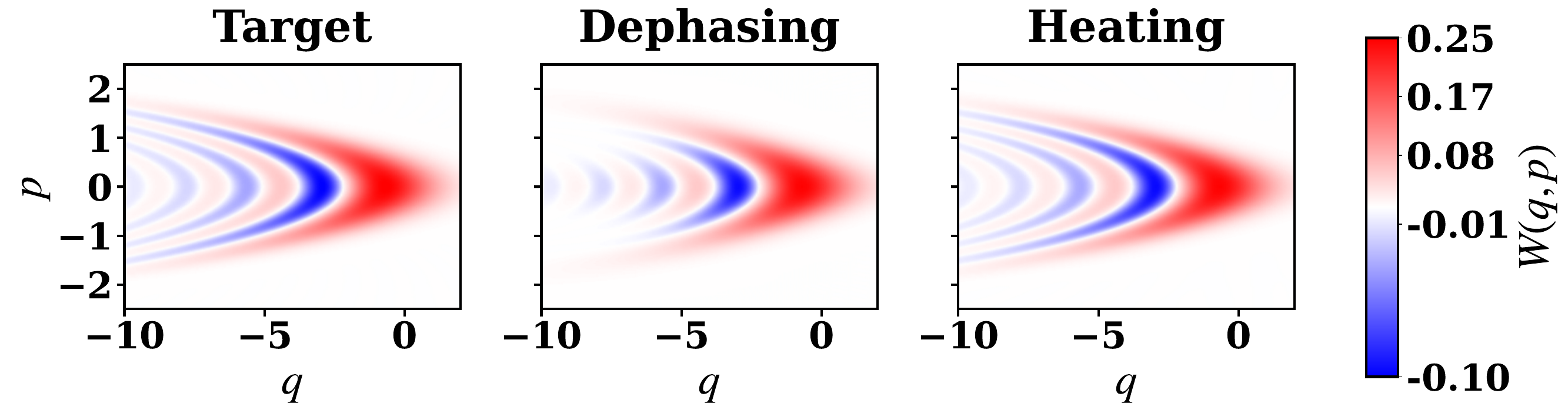}
\captionof{figure}{
Comparison of Wigner functions for the cubic phase gate acting on the vacuum state.
The target cubic phase gate on input ground state oscillator is shown together with the generated gates obtained
under motional dephasing with a coherence time $T_{coh} = 50$ ms and under heating $(\dot{n}_{th}= 10 \text{ quanta/s})$.
}
\label{fig:wigner_vacuum_target_loss_thermal}
\end{center}

\vspace{1em}

\begin{center}
\includegraphics[width=0.95\textwidth]{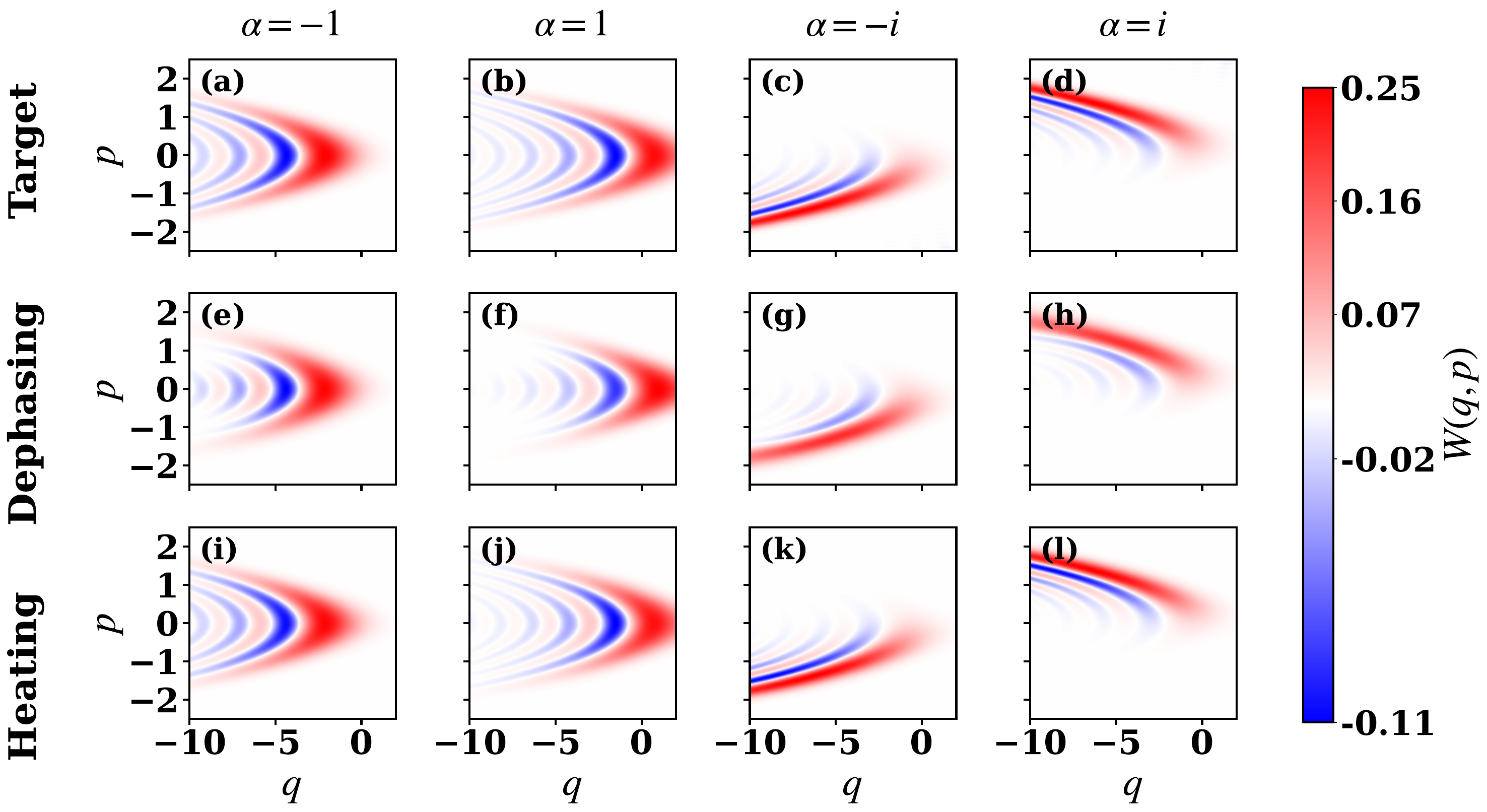}
\captionof{figure}{
Comparison of Wigner functions for target and generated cubic phase states.
For each coherent amplitude $\alpha = \{-1, 1, -i, i\}$ (from left to right),
the top row shows the target cubic phase state,
the middle row shows the generated state under motional dephasing with a coherence time $T_{coh} = 50$ ms,
and the bottom row shows the generated under heating $(\dot{n}_{th}= 10 \text{ quanta/s})$.
The comparison highlights the deformation and smoothing of phase-space negativities
induced by dissipative effects during the gate implementation.
}
\label{fig:wigner_target_loss_thermal}
\end{center}
\vspace{2em}

\begin{center}
\includegraphics[width=0.95\textwidth]{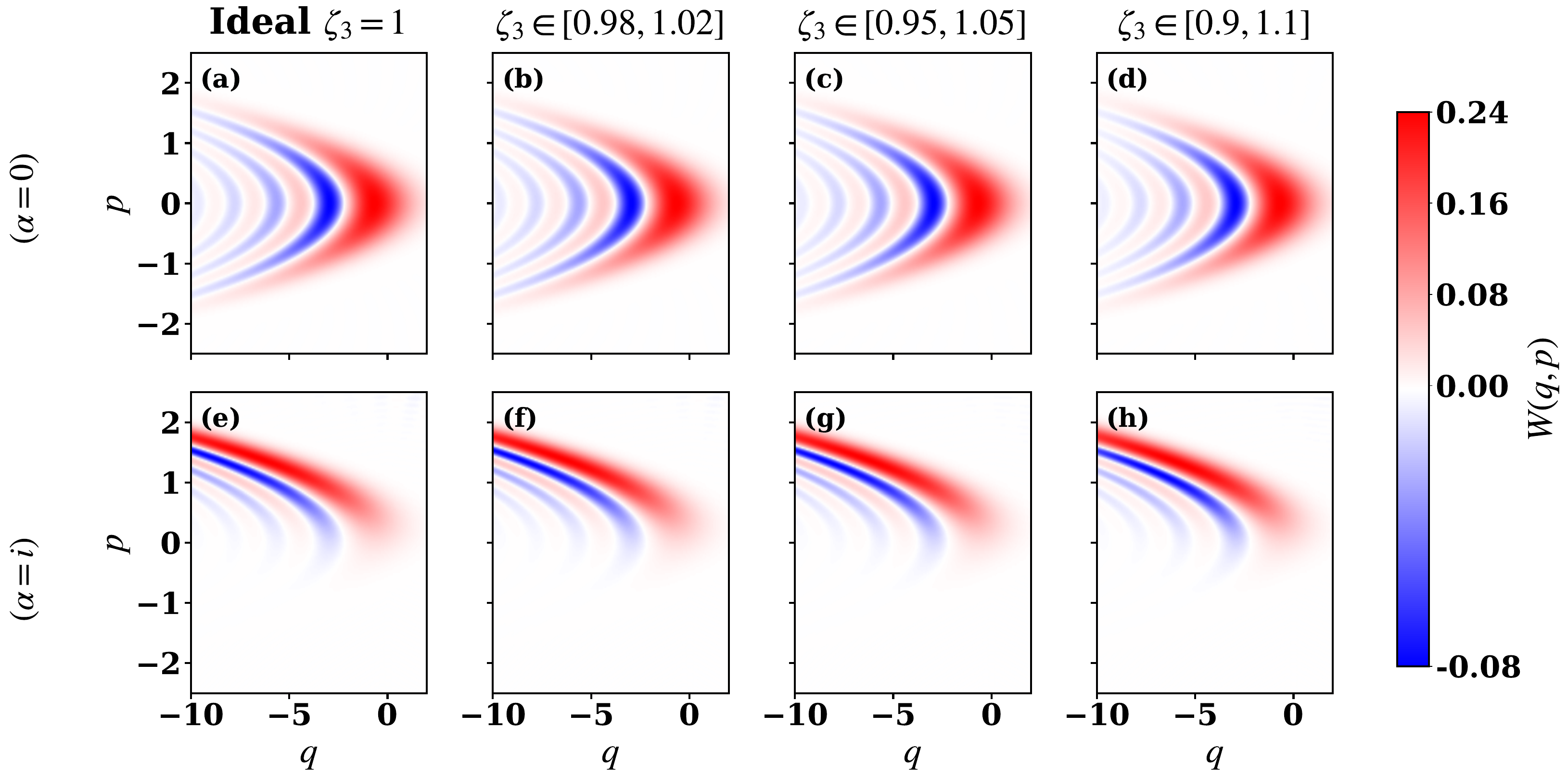}
\captionof{figure}{
Comparison between ideal and realistic cubic phase states under cubicity fluctuations.
All panels are shown using the same global color scale fixed by the ideal Wigner function.
From left to right, the columns correspond to the ideal target state and to fluctuations
of 2\%, 5\%, and 10\% in the cubicity parameter.
This representation shows that, on the scale of the ideal Wigner function,
the phase-space distributions of the generated states remain visually
indistinguishable from the target state even in the presence of cubicity
fluctuations.
To make the effect of these fluctuations explicit, a difference representation
is introduced figure \ref{fig:sm_wigner_difference}.
}
\label{fig:sm_wigner_ideal_vs_fluct}
\end{center}

\vspace{1em}

\begin{center}
\includegraphics[width=0.95\textwidth]{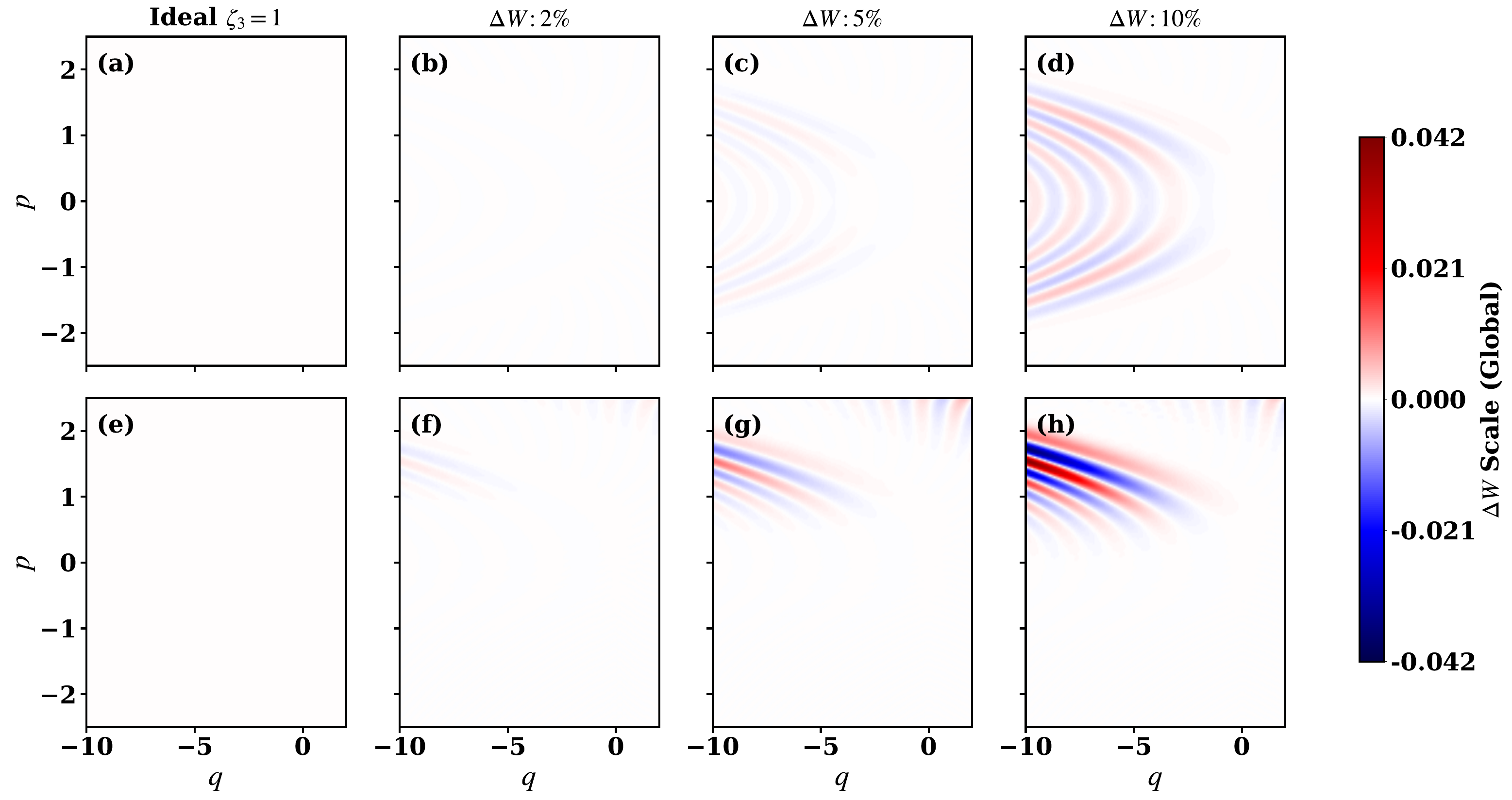}
\captionof{figure}{
Difference Wigner functions $\Delta W(q,p)=W_f(q,p)-W_i(q,p)$ between the fluctuating
cubic phase states and the ideal target state.
All difference plots use the same global color scale fixed by the ideal Wigner function.
While the absolute Wigner functions appear nearly identical for small fluctuations,
the difference representation reveals the progressive growth of phase-space deviations
as the fluctuation strength increases.
}
\label{fig:sm_wigner_difference}
\end{center}

\end{widetext}
Figure~\ref{fig:wigner_vacuum_target_loss_thermal} illustrates Wigner function deformation under dephasing and heating ($\dot{n}_{th} = 10$ quanta/s \cite{Talukdar2016}) for ground state input. Dissipation smooths interference fringes and reduces negativity depth.  This extends  to coherent states $\alpha \in \{-1, 1, -i, i\}$, showing that while noise deforms the state, the non-Gaussian crescent structure remains robust across phase space.

Figure~\ref{fig:sm_wigner_ideal_vs_fluct} evaluates the ideal gates against static cubicity fluctuations of 2\%, 5\%, and 10\%. The Wigner distributions remain visually indistinguishable from the ideal target at these scales. To resolve minor discrepancies, Figure~\ref{fig:sm_wigner_difference} plots the difference Wigner functions $\Delta W(q, p)$. These plots reveal that errors accumulate in high-frequency interference regions, confirming that infidelity at large $|\alpha|$ arises from predictable phase-space shifts rather than structural breakdown.

\section{Convergence of the Density Matrix}
\label{appendix:convergence}

\begin{figure*}[t]
  \centering
  \includegraphics[width=0.95\textwidth]{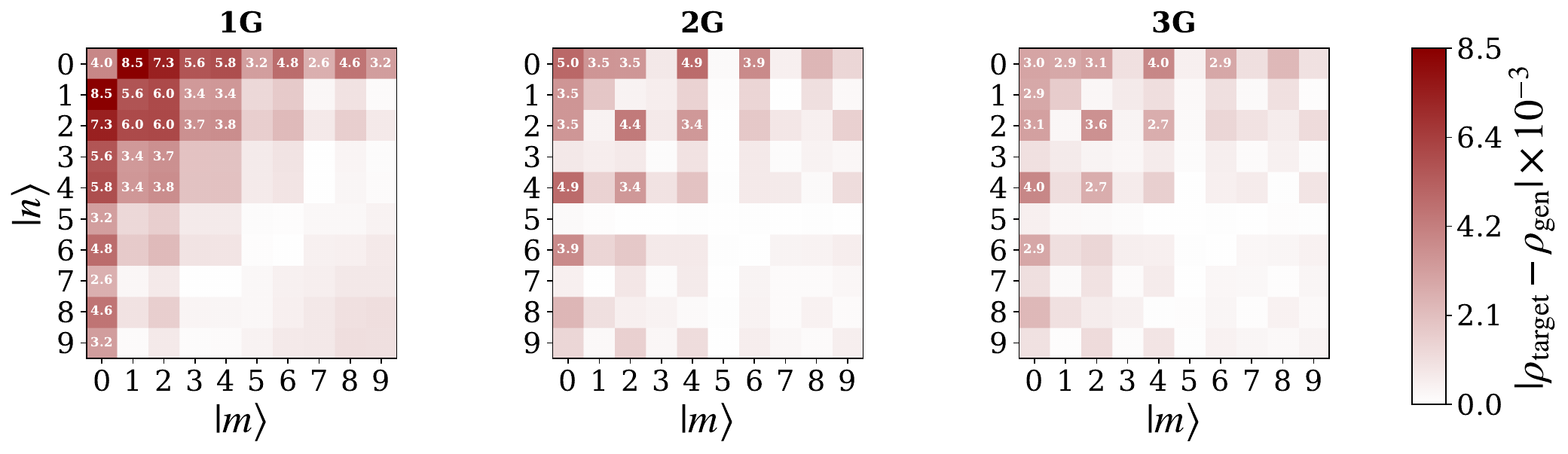}
  \caption{Convergence of the density matrix elements with increasing gate count.
  The absolute differences $|\langle m|\rho_{\rm target}-\rho_{\rm gen}|n\rangle| \times 10^{-3}$
  between target and generated states are shown for (a) 1-, (b) 2-, and (c) three-gate protocols.
  The progressive reduction in matrix element differences visually demonstrates the convergence
  toward the target cubic phase state. The color scale represents the magnitude of differences in units of $10^{-3}$.}
  \label{fig:density_convergence}
\end{figure*}

To provide a more detailed view of this convergence, Fig \ref{fig:density_convergence} displays the absolute differences between the density matrix elements of the target and generated states in the Fock basis. The heatmaps for the single-gate (1G), two-gate (2G), and three-gate (3G) protocols show a clear and systematic reduction in the  error. 
This visualizes how the repeated application of our optimized gate sequence causes the generated state $\rho_{\text{gen}}$ to converge rapidly to the target state $\rho_{\text{target}}$, confirming the effectiveness of the protocol in building up the correct quantum state structure.

\section{Derivation of Effective Cubicity}
\label{append:effcub}

The third sideband Hamiltonian $H_3$ is expanded beyond the Lamb-Dicke regime as:
\begin{equation}
H_3 \approx \frac{-i\Omega_3 \hat{\sigma}_y}{4} \left[ \underbrace{\frac{\eta^3}{6}(\hat{a}^{\dagger 3} - \hat{a}^3)}_{\text{Target } \zeta_3} + \underbrace{\frac{\eta^5}{24}(\hat{a}^{\dagger 4}\hat{a} - \hat{a}^\dagger\hat{a}^4)}_{\text{Residual Error}} \right]
\end{equation}
Using the relation $\hat{a}^{\dagger 4}\hat{a} \approx \hat{a}^{\dagger 3}\hat{n}$, and substituting the expectation value for a coherent state $\langle \hat{n} \rangle \approx |\alpha|^2$:
\begin{equation}
H_3 \approx \frac{-i\Omega_3 \hat{\sigma}_y}{4} \left[ \frac{\eta^3}{6}(\hat{a}^{\dagger 3} - \hat{a}^3) + \frac{\eta^5}{24}(\hat{a}^{\dagger 3} - \hat{a}^3)|\alpha|^2 \right]
\end{equation}
Factoring out the target cubicity $\zeta_3 \propto \eta^3$:
\begin{equation}
\zeta_3^{eff} \approx \zeta_3 \left( 1 + \frac{1}{4}\eta^2 |\alpha|^2 \right)
\end{equation}
In general, this becomes the form $\zeta_3^{eff} = \zeta_3(1 + \gamma \eta^2 |\alpha|^2)$ where $\gamma = 0.25$ is the uncompensated coefficient. In general, it can take a different value, and here, numerically we found $\gamma\approx 0.9$.

\end{document}